\documentclass[pra,aps,twocolumn,twoside,superscriptaddress]{revtex4}

\usepackage{amsmath,amsfonts,amssymb,color,graphics,graphicx,latexsym,revsymb,amsthm,url,verbatim,appendix,epstopdf}
\usepackage{algorithm}

\usepackage{algcompatible}

\usepackage{hyperref}

\usepackage{tikz}
\usetikzlibrary{decorations.shapes}
\usetikzlibrary{shapes.symbols}
\usetikzlibrary{decorations.pathmorphing}

\floatname{algorithm}{Protocol}

\newtheorem{theorem}{Theorem}
\newtheorem{corollary}[theorem]{Corollary}
\newtheorem{lemma}[theorem]{Lemma}

\theoremstyle{definition}
\newtheorem{definition}{Definition}
\newtheorem{proposition}[definition]{Proposition}

\def\squareforqed{\hbox{\rlap{$\sqcap$}$\sqcup$}}
\def\qed{\ifmmode\squareforqed\else{\unskip\nobreak\hfil
\penalty50\hskip1em\null\nobreak\hfil\squareforqed
\parfillskip=0pt\finalhyphendemerits=0\endgraf}\fi}
\def\endenv{\ifmmode\;\else{\unskip\nobreak\hfil
\penalty50\hskip1em\null\nobreak\hfil\;
\parfillskip=0pt\finalhyphendemerits=0\endgraf}\fi}
\def\Dbar{\leavevmode\lower.6ex\hbox to 0pt
{\hskip-.23ex\accent"16\hss}D}

\makeatletter
\def\bcj{\begin{conjecture}}
\def\ecj{\end{conjecture}}
\def\bcr{\begin{corollary}}
\def\ecr{\end{corollary}}
\def\bd{\begin{definition}}
\def\ed{\end{definition}}
\def\bea{\begin{eqnarray}}
\def\eea{\end{eqnarray}}
\def\bem{\begin{enumerate}}
\def\eem{\end{enumerate}}
\def\bex{\begin{example}}
\def\eex{\end{example}}
\def\bim{\begin{itemize}}
\def\eim{\end{itemize}}
\def\bl{\begin{lemma}}
\def\el{\end{lemma}}
\def\bpf{\begin{proof}}
\def\epf{\end{proof}}
\def\bpp{\begin{proposition}}
\def\epp{\end{proposition}}
\def\bqu{\begin{question}}
\def\equ{\end{question}}
\def\br{\begin{remark}}
\def\er{\end{remark}}
\def\bt{\begin{theorem}}
\def\et{\end{theorem}}

\def\btb{\begin{tabular}}
\def\etb{\end{tabular}}

\newcommand{\nc}{\newcommand}

\def\e{\epsilon}

 \nc{\bA}{{\bf A}} \nc{\bB}{{\bf B}} \nc{\bC}{{\bf C}}
 \nc{\bD}{{\bf D}} \nc{\bE}{{\bf E}} \nc{\bF}{{\bf F}}
 \nc{\bG}{{\bf G}} \nc{\bH}{{\bf H}} \nc{\bI}{{\bf I}}
 \nc{\bJ}{{\bf J}} \nc{\bK}{{\bf K}} \nc{\bL}{{\bf L}}
 \nc{\bM}{{\bf M}} \nc{\bN}{{\bf N}} \nc{\bO}{{\bf O}}
 \nc{\bP}{{\bf P}} \nc{\bQ}{{\bf Q}} \nc{\bR}{{\bf R}}
 \nc{\bS}{{\bf S}} \nc{\bT}{{\bf T}} \nc{\bU}{{\bf U}}
 \nc{\bV}{{\bf V}} \nc{\bW}{{\bf W}} \nc{\bX}{{\bf X}}
 \nc{\bZ}{{\bf Z}}

\nc{\cA}{{\cal A}} \nc{\cB}{{\cal B}} \nc{\cC}{{\cal C}}
\nc{\cD}{{\cal D}} \nc{\cE}{{\cal E}} \nc{\cF}{{\cal F}}
\nc{\cG}{{\cal G}} \nc{\cH}{{\cal H}} \nc{\cI}{{\cal I}}
\nc{\cJ}{{\cal J}} \nc{\cK}{{\cal K}} \nc{\cL}{{\cal L}}
\nc{\cM}{{\cal M}} \nc{\cN}{{\cal N}} \nc{\cO}{{\cal O}}
\nc{\cP}{{\cal P}} \nc{\cQ}{{\cal Q}} \nc{\cR}{{\cal R}}
\nc{\cS}{{\cal S}} \nc{\cT}{{\cal T}} \nc{\cU}{{\cal U}}
\nc{\cV}{{\cal V}} \nc{\cW}{{\cal W}} \nc{\cX}{{\cal X}}
\nc{\cZ}{{\cal Z}}

\nc{\hA}{{\hat{A}}} \nc{\hB}{{\hat{B}}} \nc{\hC}{{\hat{C}}}
\nc{\hD}{{\hat{D}}} \nc{\hE}{{\hat{E}}} \nc{\hF}{{\hat{F}}}
\nc{\hG}{{\hat{G}}} \nc{\hH}{{\hat{H}}} \nc{\hI}{{\hat{I}}}
\nc{\hJ}{{\hat{J}}} \nc{\hK}{{\hat{K}}} \nc{\hL}{{\hat{L}}}
\nc{\hM}{{\hat{M}}} \nc{\hN}{{\hat{N}}} \nc{\hO}{{\hat{O}}}
\nc{\hP}{{\hat{P}}} \nc{\hR}{{\hat{R}}} \nc{\hS}{{\hat{S}}}
\nc{\hT}{{\hat{T}}} \nc{\hU}{{\hat{U}}} \nc{\hV}{{\hat{V}}}
\nc{\hW}{{\hat{W}}} \nc{\hX}{{\hat{X}}} \nc{\hZ}{{\hat{Z}}}

\nc{\hn}{{\hat{n}}}

\def\max{\mathop{\rm max}}

\def\ox{\otimes}

\newcommand{\ket}[1]{|#1\rangle}

\newcommand{\tgate}{{\sf T}}
\newcommand{\hgate}{{\sf H}}
\newcommand{\pgate}{{\sf P}}

\newcommand{\cnot}{{\sf CNOT}}

\newcommand{\Xgate}{{\sf X}}
\newcommand{\Zgate}{{\sf Z}}

\begin{document}
\title{Quantum preprocessing for information-theoretic security in two-party computation}
\author{Li Yu}\email{yuli@hznu.edu.cn}
\affiliation{Department of Physics, Hangzhou Normal University, Hangzhou, Zhejiang 311121, China}

\begin{abstract}
In classical two-party computation, a trusted initializer who prepares certain initial correlations, known as one-time tables, can help make the inputs of both parties information-theoretically secure. We propose some bipartite quantum protocols with possible aborts for approximately generating such bipartite classical correlations with varying degrees of privacy, without introducing a third party. Under some weak requirements for the parties, the security level is nontrivial for use in bipartite computation. We show that the security is sometimes dependent on the noise level, but we propose a method for dealing with noise. The security is ``forced security'', which implies that the probability that some useful one-time tables are generated can approach $1$ in the noiseless case under quite weak assumptions about the parties, although the protocols allow aborts. We show how to use the generated one-time tables to achieve nontrivial information-theoretic security in generic two-party classical or quantum computation tasks, including (interactive) quantum homomorphic encryption. Our methods provide check-based implementations of some no-signaling correlations, including the PR-box type, with the help of communication which carry no information about the inputs in the generated correlations.
\end{abstract}
\maketitle


\section{Introduction}\label{sec1}

The security of two-party computation is a main research topic in classical cryptography. The goal is usually to correctly compute some function of the inputs from the two parties, while keeping the inputs as private from the opposite party as possible. This has been studied using classical homomorphic encryption techniques \cite{Gentry09,brakerski2011efficient} or through implementing Yao's ``Garbled Circuit'' solution \cite{Yao86}. Another possibility is to introduce a trusted third party, who may sometimes interact with the two parties for multiple rounds. To lower the requirement on the trusted third party, a ``trusted initializer'' has been proposed \cite{Beaver98}. Such trusted initializer only prepares some initial correlations between the two parties, and does not interact with any party afterwards. A trusted initializer who prepares certain initial correlations, referred to as ``one-time tables'', can help make the bipartite computation secure.

Secure two-party quantum computation is the corresponding problem in quantum computing and quantum cryptography. The two parties wish to correctly compute an output according to some public or private program while keeping their (quantum) inputs as secure as possible. Special cases of this general problem include quantum homomorphic encryption (QHE) \cite{rfg12,MinL13,ypf14,Tan16,Ouyang18,bj15,Dulek16,NS17,Lai17,Mahadev17,ADSS17,Newman18,TOR18}, secure assisted quantum computation \cite{Ch05,Fisher13}, computing on shared quantum secrets \cite{Ouyang17}, and physically-motivated secure computation (e.g. \cite{OTF19}). In the study of QHE, it is found that secure computation of the modulo-$2$ inner product of two bit strings provided by the two parties is a key task, and the one-time tables mentioned above turn out to be helpful for this task.

In this work, we propose two-party quantum protocols with aborts as replacements for the trusted initializer in preparing the one-time tables, and show that the prepared one-time tables can help achieve nontrivial degrees of information-theoretic security in bipartite classical or quantum computation. Our main protocols are based on Protocol~\ref{ptl:NLAND} which implements the following task with partial privacy: it takes as input two locally-generated uniformly random bits $x$ and $y$ from Alice and Bob, respectively, and outputs $(x\,\rm{AND}\,y)\,\rm{XOR}\,r$ on Alice's side and $r$ on Bob's side, where $r$ is a uniformly random bit. The one-time table contains four bits: two input bits and two output bits. By putting the possible aborts in the preprocessing which does not involve useful data, we partly avoid the problem of data leakage in those aborted runs in other possible protocols with aborts.

Security in quantum key distribution \cite{BB84} is dependent on verifications. Inspired by this, we propose some protocols that verify the correctness of Protocol~\ref{ptl:NLAND}. We propose Protocol~\ref{ptl:precompute1} to select some one-time tables generated by Protocol~\ref{ptl:NLAND}. It allows Bob to abort during the protocol when he finds that Alice is cheating. When Protocol~\ref{ptl:precompute1} is used in a generic interactive bipartite classical computation with the roles of Alice and Bob switched, the data leakage of Alice is asymptotically vanishing for noiseless physical systems, but for noisy physical systems, the leakage is linearly related to the noise level. The data privacy of Bob is partial: the leakage is about half of his input bits, but the privacy is better in the case that the function is a many-to-one map for Bob's input, including the case that the function effectively evaluates universal circuits.

We then propose Protocol~\ref{ptl:precompute1b} which includes checks from both sides to ensure that the average rate of cheating by any party is asymptotically vanishing. For the bipartite computation task, the data leakage of any party is asymptotically vanishing for noiseless systems, while for noisy systems, the leakage of both parties are linearly related to the noise level.

We then propose Protocol~\ref{ptl:precompute2} which combines several one-time tables generated by Protocol~\ref{ptl:precompute1} or \ref{ptl:precompute1b} into one. When Protocol~\ref{ptl:precompute2} based on Protocol~\ref{ptl:precompute1} is used in bipartite classical computation, the data leakage of Alice is exponentially small, so it is almost independent of the physical noise, while some polynomial overhead is needed to make the data privacy of Bob comparable to that in Protocol~\ref{ptl:precompute1}. But such polynomial overhead is not too bad, since the function to be computed can be recompiled in general, as discussed in Sec.~\ref{sec7}. To deal with noise, we propose Protocol~\ref{ptl:precompute3} which has enhanced correctness. It detects the errors (from noise and possible malicious activity) with some good chance, such that the error rate in the output is polynomially small, while the security level is similar to that of Protocol~\ref{ptl:precompute1b} at the cost of a polynomial-factor increase in resources. It also has a variant of combining one-time tables in the similar way as in Protocol~\ref{ptl:precompute2}. The resource overhead in Protocol~\ref{ptl:precompute3} is exponentially large if the output error rate is required to be exponentially small. However, we think that polynomially small error in the output is sometimes acceptable, since the circuit to be evaluated is usually of polynomial length.

All the protocols are secure if both parties are honest-but-curious. An honest-but-curious party is one who follows the protocol while possibly making measurements which do not affect the final computation result. In our protocols, an honest-but-curious party does not learn anything about the other party's data, no matter whether the other party cheats or not.

In our protocols, when one party is honest-but-curious and the other party is malicious, the privacy of the data of the honest-but-curious party is guaranteed to reach the targeted level even if the other party cheats. In Protocol~\ref{ptl:precompute1}, we assume that Bob is \emph{conservative}, meaning that he values the privacy of his data higher than the possibility to learn Alice's data. Operationally, this means he always performs the checking and aborts when the number of wrong instances exceeds the threshold that he had chosen. Alice needs to be \emph{weakly cooperating} for the protocols not to abort, meaning that she does not cheat much in some batch of the instances of Protocol~\ref{ptl:NLAND}. For Alice's data security to be enhanced by her verifications in Protocol~\ref{ptl:precompute1b}, she should be conservative in the sense described above. But partly due to the possible aborts, it actually suffices to assume one of the parties is conservative in Protocol~\ref{ptl:precompute1b}, since then the other party might as well be conservative to reach a better security level for himself (herself). Although Protocol~\ref{ptl:precompute2} is quite effective when there is no noise (including errors), it may not be better than Protocol~\ref{ptl:precompute1} or \ref{ptl:precompute1b} when there is some non-negligible level of noise. In the noisy case, we propose using Protocol~\ref{ptl:precompute1b} or Protocol~\ref{ptl:precompute3}, where the latter is for performing some error detection while not harming security too much.

The security of the protocols is ``forced security'', which means Alice is forced by Bob's checks to not cheat in some batches of Protocol~\ref{ptl:NLAND}. It implies that the probability that some one-time tables with targeted (partial) security are generated would approach $1$ in the noiseless case under quite weak assumptions about the parties (that Alice weakly cooperates by not cheating in some batches of Protocol~\ref{ptl:NLAND}, and Bob indeed does the checks due to that he is conservative), although the protocols allow aborts.

We show some applications in general two-party classical computation, and the check-based implementations of 1-out-of-2 oblivious transfer and bit commitment under some assumptions mentioned above. To enjoy some quantum speedup together with the security benefit brought about by our preprocessing, we propose an interactive QHE scheme with costs polynomial in circuit size, as well as a constant-round QHE scheme with exponential cost, which use the precomputed one-time tables as a resource, but both schemes have more rounds of communication than in the original definition of QHE. Such scheme is then generalized to general two-party quantum computation with a publicly known circuit and private inputs on both parties, and to the case of private circuit provided by one party and private inputs on both parties. Our protocols provide check-based implementations of some no-signaling correlations with the help of classical communication which do not carry information about the inputs in the generated correlations.

The rest of the paper is organized as follows. Sec.~\ref{sec2} contains some introduction of the background. In Sec.~\ref{sec3} we introduce the quantum protocols for generating the one-time tables. Sec.~\ref{sec4} shows applications in general two-party classical computation. Sec.~\ref{sec5} shows applications in general two-party quantum computation. Sec.~\ref{sec6} shows applications in check-based implementations of some no-signaling correlations with the help of classical communication. Sec.~\ref{sec7} contains some discussions about the security in the noisy case, and physical implementations. Sec.~\ref{sec8} contains the conclusion and some open problems.

\section{Preliminaries}\label{sec2}

On computing two-party classical functions with quantum circuits, Lo \cite{Lo97} studied the data privacy for publicly known classical functions with the output on one party only. Buhrman \emph{et al} \cite{bcs12} studied the security of two-party quantum computation for publicly known classical functions in the case that both parties know the outcome, although with some limitations in the security notions. These and other results in the literature \cite{Colbeck07} suggest that secure bipartite classical computing cannot be generally done by quantum protocols where the two parties have full quantum capabilities. In the current work, the protocols allow aborts in the quantum preprocessing (Bob may abort when he detects that Alice has cheated), and local randomness is used, so the scenario considered here does not fit into the assumptions in the works mentioned above. We assume that one party values the privacy of his data higher than the possibility to learn the other party's data. Under such assumption, we do not require the parties in the main bipartite computation stage to be entirely classical.

Next, we introduce the simplest case in the one-time tables \cite{Beaver98}. Actually the type of table discussed below is known as precomputed oblivious transfer, although our usage of such table is not in the form of transferring a bit. Rather,  it is more like implementing a gate. The bipartite AND gate with distributed output is a gate that takes as input two distant bits $a$ and $b$, and outputs $(a\cdot b)\oplus r$ and $r$ on the two parties, respectively, where $r$ is a uniformly random bit. (XOR is denoted as $\oplus$; AND is denoted as the $\cdot$ symbol.) It is sufficient for secure two-party classical computation, although there may be other constructions. Theoretically, the bipartite AND gate with distributed output on two distant input bits $a$ and $b$ can be computed while keeping both input bits completely private, with the help of a precomputed ideal one-time table of the nonlocal-AND type. Such one-time table has two locally-generated uniformly random bits $x$ and $y$ on Alice's and Bob's side, respectively, and also has $(x\cdot y)\oplus r$ and $r$ on Alice's and Bob's side, respectively, where $r$ is a uniformly random bit. The steps for the bipartite AND-gate computation with distributed output are as follows:

1. Alice announces $a'=a \oplus x$. Bob announces $b'=b \oplus y$.

2. Each party calculates an output bit according to the one-time table and the received message. Alice's output is $(x\cdot b')\oplus(x\cdot y)\oplus r$. Bob's output is $(a'\cdot b)\oplus r$.

The XOR of the two output bits is $(x\cdot b')\oplus(x\cdot y)\oplus r\oplus(a'\cdot b)\oplus r=a \cdot b$, while each output bit is a uniformly random bit when viewed alone, because $r$ is a uniformly random bit. Since the messages $a'$ and $b'$ do not contain any information about $a$ and $b$, the desired bipartite AND gate is implemented while $a$ and $b$ are still perfectly private.

Some notations are as follows.  By ``forced security'', we mean that the security in a protocol is guaranteed by verifications where failure to pass them would cause the protocol to abort. By saying that a protocol is ``cheat-sensitive'', we mean that any cheating will probably cause the protocol to abort. Our protocols starting from Protocol~\ref{ptl:precompute1} are in fact cheat-sensitive, but to avoid confusion with protocols in the literature (e.g. \cite{HK04}) which have different levels of security and effectiveness from ours, we do not use the term in the titles of the protocols.

Denote $\ket{\tilde 0}=\ket{+}:=\frac{1}{\sqrt{2}}(\ket{0}+\ket{1})$, and $\ket{\tilde 1}=\ket{-}:=\frac{1}{\sqrt{2}}(\ket{0}-\ket{1})$. The Hadamard gate $\hgate$ satisfies $\hgate\ket{0}=\ket{+}$ and $\hgate\ket{1}=\ket{-}$. The random bits are unbiased and independent of other variables by default. An EPR pair is two qubits in the state $\frac{1}{\sqrt{2}}(\ket{00}+\ket{11})$.

\section{The quantum protocols for generating one-time tables}\label{sec3}

The main quantum protocols to be introduced later are based on Protocol~\ref{ptl:NLAND}, which is the revised version of a subprocedure of a protocol from \cite{Yu18}. The Protocol~\ref{ptl:NLAND} effectively computes an AND function on two remote classical bits from the two parties, with the output being a distributed bit, i.e. the XOR of two bits on the two parties. The security is not ideal: the plain use of such protocol would give rise to non-ideal security in (interactive) quantum homomorphic encryption \cite{Yu18}, and the security is such that some additional verification need to be added in the protocol for it to be nontrivial. Later we propose  protocols that check and sometimes combine the one-time tables generated from Protocol~\ref{ptl:NLAND}, to be used as a preprocessing stage for a bipartite classical or quantum computation task.

The Protocol~\ref{ptl:NLAND} involves direct sending of states, while the Protocol~\ref{ptl:NLAND2} in Appendix~\ref{appent} is the corresponding entanglement-based variant. The Protocol~\ref{ptl:NLAND2} uses prior shared entanglement to remotely prepare some state on Bob's side via Alice's local measurements, and it also involves a step of teleportation \cite{bbc93} from Bob to Alice with partial information about the corrections withheld by the sending party. The teleportation approach allows Alice and Bob to do operations simultaneously, see the discussions in Secs.~\ref{sec6} and \ref{sec7}. These two protocols crucially depend on the property of the $\cnot$ gate: it is equivalent to a $\cnot$ gate in the reverse direction in an unbiased basis (the $X$ basis on both qubits).

\begin{algorithm*}[htb]
\caption{A two-message quantum protocol for generating one-time tables with partial privacy}\label{ptl:NLAND}
\begin{flushleft}
\noindent{\bf Input:} A random bit $x$ from Alice and a random bit $y$ from Bob.\\
\noindent{\bf Output:} $(x\cdot y)\oplus r$ on Alice's side, and $r$ on Bob's side, where $r$ is a random bit.\\
\noindent The input and output together form the one-time table.\\
\begin{enumerate}
\item Alice generates two random bits $s$ and $t$. If $s=0$, she prepares the state $\ket{x}\ket{t}$; if $s=1$, she prepares the state $\ket{\tilde t}\ket{\tilde x}$, where the tilde represents $X$-basis encoding. She sends the prepared two-qubit state to Bob.
\item Bob receives the two qubits. If $y=0$, Bob does a $\cnot$ gate on the two qubits, with the first qubit being the control qubit.
\item He generates two random bits $h_1$ and $h_2$. He does a $\sigma_y$ gate on any qubit where the corresponding bit $h_j$ ($j=1,2$) is $1$. He generates a random bit $p$. If $p=1$, he does $\sigma_z$ gates on both qubits. He sends the two qubits to Alice. The bit $h:=h_1 \oplus h_2$ is his output.
\item Alice receives the two qubits. If $s=0$, Alice measures the two received qubits in the $Z$ basis, otherwise she measures them in the $X$ basis. She calculates the XOR of three bits: the two outcome bits, and the $t$. The obtained bit is her output.
\end{enumerate}
\end{flushleft}
\end{algorithm*}

In Protocol~\ref{ptl:NLAND}, Alice's input bit has partial privacy even for a cheating Bob, while Bob's input bit is secure for an honest-but-curious Alice, but is not secure at all for a cheating Alice. The privacy of Alice's input bit $x$ can be quantified using the accessible information or the trace distance. The accessible information, i.e. the maximum classical mutual information corresponding to Bob's possible knowledge about Alice's input, is exactly $\frac{1}{2}$ bits, which happens to be equal to the Holevo bound in the current case. For a cheating Bob to get the maximum amount of information, his best measurement strategy in the current case is to use a fixed projective measurement: to measure the first qubit in the $Z$ basis, and the second qubit in the $X$ basis. The trace distance of the two density operators for
Alice's two possible input values is $\frac{1}{2}$, by direct calculation. Thus, the probability that Bob guesses Alice's input bit correctly is $(1+\frac{1}{2})\cdot\frac{1}{2}=\frac{3}{4}$. Note that with this particular measurement just mentioned, he cannot make the distributed output of the one-time table correct. In other words, Bob cannot learn the other party's input without consequences.

To learn about Bob's input bit, a cheating Alice may use an entangled state $\frac{1}{2}(\ket{00}+\ket{01}+\ket{10}-\ket{11})$. From Bob's returned state, Alice may find out Bob's input bit with certainty. But in such case Alice has no effective input to speak of, and she does not know Bob's output bit $h$, so even if she chooses an input bit for herself later, she cannot determine her output bit for making the distributed output correct.

The entanglement-based version for Protocol~\ref{ptl:NLAND} is Protocol~\ref{ptl:NLAND2} in Appendix~\ref{appent}. Note that in Protocols~\ref{ptl:NLAND} and \ref{ptl:NLAND2}, Alice may cheat by declaring that she did not receive some qubits in some instances. In such case she may be certain about one of the three bits: $y$, $y\oplus r$ or $r$, depending on her measurement strategy, but she cannot learn both $y$ and one of $r$ and $y\oplus r$ (near) perfectly. Since this is a somewhat non-conventional way of cheating (as there is significant ratio of failed instances noticeable to Bob), we discuss this case near the end of Appendix~\ref{appent}.

The Protocol~\ref{ptl:NLAND} has two stages of communication. The following Protocol~\ref{ptl:NLAND3} has only one stage of communication. It is derived from  the entanglement-based Protocol~\ref{ptl:NLAND2}. Instead of using prior entanglement, Bob prepares some pure state on four qubits dependent on his private keys and a private input bit $y$. Bob sends such state to Alice. Since there are some different choices of such pure state, Alice receives a mixed state in her view. She does some single-qubit measurements in some bases of her choice to obtain some classical correlation (one-time table) shared with Bob.

\begin{algorithm*}[htb]
\caption{A one-message quantum protocol for generating one-time tables with partial privacy}\label{ptl:NLAND3}
\begin{flushleft}
\noindent{\bf Input of the generated one-time table:} A random bit $x$ generated in the protocol, and a random bit $y$ from Bob.\\
\noindent{\bf Output of the generated one-time table:} $(x\cdot y)\oplus r$ on Alice's side, and $r$ on Bob's side, where $r$ is a random bit.\\
\noindent The input and output together form the one-time table.\\
\begin{enumerate}
\item Bob initializes four qubits in a computational basis state $\ket{i_1}\ket{i_2}\ket{i_3}\ket{i_4}$, where the choice of $i_1,i_2,i_3$ are random, and $i_4=i_1 \oplus i_2 \oplus i_3$. He denotes $r:=i_1\oplus i_2$. He performs the Hadamard gate on the first two qubits. He then performs the gates $\cnot_{13}$ and $\cnot_{24}$, where the subscripts are labels for the qubits, and the first label is for the controlling qubit. If $y=0$, he does a $\cnot_{12}$ gate. He sends the four qubits to Alice. Bob's output bit is $r$.
\item Alice receives the four qubits from Bob. She generates a random bit $s$. If $s=0$, she measures the four qubits in the $Z$ basis, and records the measurement outcome on the first qubit as $x$; if $s=1$, she measures the four qubits in the $X$ basis, and records her measurement outcome on the second qubit as $x$. (Under $X$-basis measurement, the states $\ket{+}$ and $\ket{-}$ are recorded as $0$ and $1$, respectively.) The XOR of the measurement outcomes on the three remaining qubits is recorded as $g$, which is regarded as Alice's output bit.
\end{enumerate}
\end{flushleft}
\end{algorithm*}

In Protocol~\ref{ptl:NLAND3}, Bob's input $y$ is partly secure. For an honest Bob, Alice can know Bob's input $y$ with probability $\frac{3}{4}$, since the two density operators of Bob's have trace distance $\frac{1}{2}$. On the other hand, Alice's input in the one-time table, $x$, is completely insecure in the worst case for a cheating Bob, but it is completely secure for an honest Bob. A cheating Bob could prepare some state such that the $x$ is determined, such as preparing $\ket{0}\ket{+}$ on the first two qubits, but then the output of the one-time table would be completely random (due to that Alice would measure one of the first two qubits with an unbiased basis compared to the basis for the state of the qubit), hence it is incorrect with large probability. Overall, the security characteristics of Protocol~\ref{ptl:NLAND3} is analogous to that of Protocol~\ref{ptl:NLAND} but with the roles of the two parties switched. But the fewer rounds of communication makes Protocol~\ref{ptl:NLAND3} potentially interesting in other applications. In later parts of the paper, we take Protocol~\ref{ptl:NLAND} as the example in analyzing some composed protocols.

In the following we present protocols which check or combine the one-time tables generated in Protocol~\ref{ptl:NLAND}. The first one has partial security for Alice and near-perfect security for Bob, while the second one involves checking by both parties, and aims for near-perfect security for both parties. The third one aims for near-perfect security for both parties with emphasis on the security of one party.

\begin{algorithm*}[htb]
\caption{A partly-secure protocol for checking the one-time tables}\label{ptl:precompute1}
\begin{enumerate}
\item Alice and Bob perform many instances of Protocol~\ref{ptl:NLAND} (sequentially or in parallel) to generate some one-time tables, and exchange messages to agree on which instances were successfully implemented experimentally. Suppose $m$ one-time tables were implemented. The one-time tables labeled by $j$ has inputs $a_j$ and $b_j$, and outputs $e_j$ and $f_j$.
\item Bob randomly selects $K$ integers in $\{1,\cdots,m\}$, which are labels for which one-time table. He tells his choices to Alice. The integer $K$ satisfies that $m-K$ is an upper bound on the number of required one-time tables in the main bipartite computing task, and the ratio $\frac{K}{m}$ is related to the targeted security level of the overall computation.
\item Alice sends the bits $a_j$ and $e_j$ to Bob for all chosen labels $j$.
\item For any chosen label $j$, Bob checks whether $a_j$ and $e_j$ satisfy that $a_j \cdot b_j=e_j \oplus f_j$. If the total number of failures is larger than some preset number of Bob's (e.g. $0$, or a small constant times $K$), he aborts the protocol, or restarts the protocol to do testing on a new batch of instances of Protocol~\ref{ptl:NLAND} if the two parties still want to perform some secure two-party computation. Otherwise, the remaining one-time tables are regarded as having passed the checking and will be used later in the two-party computing task. They may repeat the steps above to prepare more one-time tables on demand.
\end{enumerate}
\end{algorithm*}

In Protocol~\ref{ptl:precompute1}, Alice's input bit has partial privacy, which is the same as in the analysis of Protocol~\ref{ptl:NLAND} above. When the ratio $\frac{K}{m}$ is near one, the nonlocal correlations in the remaining unchecked one-time tables can be regarded as almost surely correct. This is because of Bob's checking. We require Alice to be weakly cooperating, that is, she does not cheat in some of the batches of instances, since otherwise no one-time table may pass the test.  Some degree of weak cooperation is required for two parties to perform a computation anyway, and the above assumption of Alice has no effect on the data security of any party when Bob satisfies the assumption below, thus we may ignore the assumption above and just state the following assumption on Bob as the requirement of our protocols. In the following we assume that Bob is conservative, which means that he values the privacy of his data higher than the possibility to learn Alice's data. Later in Sec.~\ref{sec4} we will see that it effectively implies that he indeed does the checking. For an honest-but-curious Alice, the resulting correlation is correct, and she does not learn anything about Bob's input bit $y$ (using the notations in Protocol~\ref{ptl:NLAND}, same below). In the following we discuss the case that Alice cheats.

If Alice cheats and gets at least partial information about Bob's input bit $y$, the state sent from Alice to Bob must be different from what is specified in the protocol; her best choice of state for cheating is mentioned previously. To pass Bob's test while learning about Bob's input $y$, she should know both $y$ and $r$, or know both $y$ and $y \oplus r$. (The two conditions are equivalent in the exact case, but not necessarily equivalent in the partial-information case.)  In the following, let $I^{\cal M}_y$ denote the classical mutual information learnable by Alice about Bob's bit $y$ (with uniform prior distribution) if she uses the measurement $\cal M$ on the received two qubits (possibly a POVM measurement), in an instance of Protocol~\ref{ptl:NLAND}. The $I^{\cal M}_r$ and $I^{\cal M}_{y\oplus r}$ are defined similarly, but note that they are conditioned on the uniform distribution for $y$, similar to the case of $I^{\cal M}_y$.

\bpp\label{prop1}
In Protocol~\ref{ptl:NLAND}, the following inequalities hold:
\bea
I^{\cal M}_y+I^{\cal M}_r &\le& 1,\label{eq:info1}\\
I^{\cal M}_y+I^{\cal M}_{y\oplus r} &\le& 1,\label{eq:info2}\\
I^{\cal M}_y+\max(I^{\cal M}_r,I^{\cal M}_{y\oplus r}) &\le& 1.\label{eq:info3}
\eea
where the two $\cal M$ are the same in each equation. All the quantities on the left-hand-sides are also dependent on Bob's received state $\sigma_A$. It is effectively prepared by Alice, and is a mixed state on two qubits (in numerical calculations, it is viewed as a pure state on two of Bob's qubits and two imaginary ancillary qubits), and the two $\sigma_A$ are the same in each equation. We abbreviate the symbol $\sigma_A$.
\epp

Note that the relationship between $\sigma_A$ and $x$ is as follows: if Alice is honest, the $\sigma_A$ is determined by the choice of $x$ according to Protocol~\ref{ptl:NLAND} up to some Pauli operators (arising from the teleportations in Step 2 of Protocol~\ref{ptl:NLAND}). If Alice is dishonest, the $\sigma_A$ is not necessarily related to $x$ (since the latter may be undefined), and it may be a mixed state in Bob's view, but Alice may hold the purification for it, where the purification system needs to include two ancillary qubits at most. In defining $\sigma_A$, we use Bob's received state instead of Alice's input state before teleportation, since it is more general: Alice could cheat by changing her operations to deviate from the original operations in the teleportation, but she always effectively prepares a (mixed) state on Bob's two qubits no matter what she does.

\bpf
Suppose $\sigma_A$ is Bob's received two-qubit mixed state. The overall communication from Bob to Alice in Protocol~\ref{ptl:NLAND} is effectively only one classical bit, since if Bob randomly performs a $\sigma_z$ gate on his first sent qubit, the sent two qubits would be in a maximally mixed state, containing no information for Alice. Also note that there are effectively no other prior correlations between the two parties besides the fixed entangled state, so the locking of information \cite{DHL04} does not occur here. The amount of information that Alice learns about the joint distribution of $y$ and $r$ is upper bounded by $1$ bit. The bits $y$ and $r$ are independent when Bob produces them, so the $y$ and $r$ are independent prior to Alice's measurement. Thus the inequality~\eqref{eq:info1} holds, where we have assumed that the two $\sigma_A$ implicit in the information quantities are the same in this equation (same below). The bits $y$ and $y\oplus r$ jointly determine $y$ and $r$, and vice versa, so the amount of information that Alice learns about the joint distribution of $y$ and $y\oplus r$ is upper bounded by $1$ bit. And since the bits $y$ and $y\oplus r$ are independent prior to Alice's measurement, we have that the inequality~\eqref{eq:info2} holds. The inequalities \eqref{eq:info1} and \eqref{eq:info2} together imply \eqref{eq:info3}.
\epf

The probability that Alice passes Bob's test at a particular instance is related to the $\max(I^{\cal M}_r,I^{\cal M}_{y\oplus r})$ in Eq.~\eqref{eq:info3}. When the probability of passing approaches $1$, such maximum approaches $1$, then it must be that one of them approaches $1$. Then, Prop.~\ref{prop1} implies that Alice can learn almost nothing about $y$ if she measured in the same basis, but in fact a cheating Alice knows which instances are remaining and will not be checked later (although it is conceivable that some checks may be done after the main computation, see Sec.~\ref{sec7} below), so she can choose to do any measurement on the received states in these remaining instances. Such measurement may not be the same as $\cal M$ in the other term in Eq.~\eqref{eq:info3}. This implies that Eq.~\eqref{eq:info3} alone is not sufficient for proving the security of Protocol~\ref{ptl:precompute1}.

\bt\label{thm1}
In Protocol~\ref{ptl:precompute1}, Bob's input is asymptotically secure.
\et

\bpf
We first consider the case that Alice's operations are independent among different instances of Protocol~\ref{ptl:NLAND}, and at last comment that the non-independent case still satisfy the extreme case of the inequalities above, giving rise to the security of Protocol~\ref{ptl:precompute1}.

Due to the freedom of measurement basis choice mentioned above, the Holevo bounds, which are upper bounds of the information quantities, are more relevant for proving the security of Protocol~\ref{ptl:precompute1}. Under the condition that Alice's operations are independent among the instances, we need only consider the Holevo bounds for a single instance of  Protocol~\ref{ptl:NLAND}. Let $\chi_y$ be the Holevo quantity which is the upper bound for $I^{\cal M}_y$. It is defined as
\bea\label{eq:defHolevo}
\chi_y=S(\rho)-\frac{1}{2}\sum_{j=1}^2 S(\rho_j),
\eea
where $\rho_j$ is the density operator that Alice receives from Bob for the case of $y=j$ after Pauli corrections determined by Bob's sent bit, and $\rho=\frac{1}{2}(\rho_1+\rho_2)$. The $S$ represents the von Neumann entropy. The definition of $\chi_y$ shows that it is conditioned on the uniform prior distribution for $y$. The quantities $\chi_r$ and $\chi_{y\oplus r}$ are defined similarly and are also conditioned on the uniform prior distribution for $y$. We claim that the following inequality holds for small positive $\e$ and a nonnegative continuous function $f(\e)$,
\bea\label{eq:holevo4}
\chi_y+\max(\chi_r,\chi_{y\oplus r}) \le 1+f(\e),\,\quad\,\notag\\
\mbox{for}\,\,\max(\chi_r,\chi_{y\oplus r})\ge 1-\e,\notag\\
\mbox{where}\,\,f\,\,\mbox{is continuous and}\,\,f(0)=0.
\eea
The reason is as follows. The Holevo quantities in Eq.~\eqref{eq:holevo4} satisfy uniform continuity, because of the combination of the following two reasons: the ancilla in Alice's initial state $\sigma_A$ (introduced in Prop.~\ref{prop1}) is effectively at most $4$ dimensions due to the Schmidt decomposition, and note that such ancilla is also the ancilla for Alice's final state; the Holevo quantity $\chi_y$ in \eqref{eq:defHolevo} is continuous as a function of $\rho_1$ and $\rho_2$ and is therefore a continuous function of Alice's initial state $\sigma_A$, and similarly, the Holevo quantities $\chi_r$ and $\chi_{y\oplus r}$ are also continuous functions of Alice's initial state $\sigma_A$. Given that the Holevo quantities satisfy uniform continuity, we obtain Eq.~\eqref{eq:holevo4} by noting the fact that
\bea\label{eq:holevoimplication}
\max(\chi_r,\chi_{y\oplus r})=1 \Longrightarrow \chi_y=0,
\eea
where Eq.~\eqref{eq:holevoimplication} holds because $\max(\chi_r,\chi_{y\oplus r})=1$ implies that $\max(I^{\cal M}_r,I^{\cal M}_{y\oplus r})=1$ for some $\cal M$, and the latter implies $\chi_y=0$ due to the following argument: suppose $I^{\cal M}_r=1$ (the case that $I^{\cal M}_{y\oplus r}=1$ is similar), and consider the four density operators on Alice's side corresponding to four different combinations of $y$ and $r$,  then the two pairs corresponding to different $r$ must be orthogonal across the pairs. Then if the states in one pair are partially distinguishable, the left-hand-side of \eqref{eq:info3} would be greater than $1$ for some $\cal M$, which violates Prop~\ref{prop1}. The above arguments shows that $\max(I^{\cal M}_r,I^{\cal M}_{y\oplus r})=1$ for some $\cal M$ implies $\chi_y=0$, hence Eq.~\eqref{eq:holevoimplication} holds.

Alice may cheat in some instances of Protocol~\ref{ptl:NLAND} so we may define a rate of cheating. Partial cheating in a instance is converted into a fractional number of cheating instances in calculating such rate. Alice's cheating probabilities among different instances may be correlated, but that does not affect the following argument since Bob randomly chooses which instances to check. It is sort of subjective for Bob to determine the average rate of cheating from the number of wrong results and the total number of tests in Protocol~\ref{ptl:precompute1}, since it depends on the \emph{a priori} knowledge about the probability distribution for Alice's average rate of cheating, and also depends on the correlations between rates of cheating among different instances of Protocol~\ref{ptl:NLAND}. Suppose that after some checking, Bob estimates that Alice's average rate of cheating is $\e$, which is a small positive constant near $0$, then the following estimate holds for the uniform distribution of $y$ and $r$ (the uniform distribution of $y$ can be imposed by Bob since he wants to make Alice's cheating be detected, and the $r$ has uniform distribution according to Protocol~\ref{ptl:NLAND}): $\max(\chi_r,\chi_{y\oplus r})\ge 1-\e$. Hence, $\chi_y\le \e+f(\e)$ according to Eq.~\eqref{eq:holevo4}. This shows that the expected amount of information about $y$ learnable by a cheating Alice in the remaining instances of Protocol~\ref{ptl:NLAND} is arbitrarily near zero for sufficiently small $\e$, even if she measures in different bases from those for the tested instances. The word ``expected'' means that even if $L\e<1$, where $L$ is the total number of one-time tables to be used for the main computation, Alice may sometimes learn about one or a few bits of Bob's input by chance, but on average, she learns not more than $L\e$ bits. Since the information about $y$ is linearly related to the information learnable by Alice in the later main computation stage (see the bipartite AND-gate computation method in Sec.~\ref{sec2}), this shows the security of Protocol~\ref{ptl:precompute1} in the case that Alice's operations are independent among instances of Protocol~\ref{ptl:NLAND}.

In the following we consider the general case that Alice's operations are not necessarily independent among instances of Protocol~\ref{ptl:NLAND}. If Alice initially prepares some correlated quantum states among $m$ instances, the generalization of Eq.~\eqref{eq:holevoimplication} should hold, due to the similar reason as that after Eq.~\eqref{eq:holevoimplication}. Then the generalization of Eq.~\eqref{eq:holevo4} for the corresponding Holevo bounds should hold approximately near such extreme point, due to the uniform continuity of the Holevo bounds (as functions of the joint state on Bob's side on multiple subsystems). Since Bob's variables $y$ and $r$ are independent among the instances, the generalizations of Eq.~\eqref{eq:holevo4} just mentioned have the same scaling near the extreme point (as the number of instances of Protocol~\ref{ptl:NLAND} grows) as in the case that Alice's operations are independent. The last point can be seen from that Alice's states in other instances of Protocol~\ref{ptl:NLAND} serve as auxiliary systems in considering Holevo quantities of the form \eqref{eq:defHolevo}, so the one-copy tradeoff curve of the Holevo quantities still holds, i.e. Eq.~\eqref{eq:holevo4} for one instance still holds with the same quantitative levels (including near the extreme point). This shows that the argument for the security for the case of independent operations of Alice can be extended to the general case.

Finally we consider the ``restarts'' of the protocol mentioned in the end of Protocol~\ref{ptl:precompute1}. Since Bob's inputs among different runs are independent, Alice has no way of using joint initial states or making joint measurements to take advantage of the possibility of restarts. Hence the probability that a cheating Bob would pass Alice's test adds up at most additively. Similar statement can be said for Alice's cheating. And since practically there can only be a polynomial number of restarts, due to resource constraints, either party can set appropriate thresholds in his or her checking to make the overall probability of cheater passing the tests upper bounded by any small positive constant.
\epf

Some numerical results are in Appendix~\ref{app1}.

To improve Alice's security in the protocol above, we propose the following Protocol~\ref{ptl:precompute1b}, in which Alice also does some checking about Bob's behavior.

\begin{algorithm*}[htb]
\caption{A protocol for checking the one-time tables by both parties}\label{ptl:precompute1b}
\begin{enumerate}
\item Alice and Bob perform many instances of Protocol~\ref{ptl:NLAND} to generate some one-time tables, and exchange messages to agree on which instances were successfully implemented experimentally. Suppose $m$ one-time tables were implemented. The one-time tables labeled by $j$ has inputs $a_j$ and $b_j$, and outputs $e_j$ and $f_j$.
\item (The steps 2 to 4 can be done concurrently with the steps 5 to 7.) Bob randomly selects $K_B$ integers in $\{1,\cdots,m\}$, which are labels for which one-time table. He tells his choices to Alice.
\item Alice sends the bits $a_j$ and $e_j$ to Bob for all chosen labels $j$.
\item For any chosen label $j$, Bob checks whether $a_j$ and $e_j$ satisfy that $a_j \cdot b_j=e_j \oplus f_j$. If the total number of failures is larger than some preset number of Bob's (e.g. $0$, or a small constant times $m$), he aborts the protocol, or asks Alice to restart the protocol to do testing on a new batch of instances of Protocol~\ref{ptl:NLAND} if the two parties still want to perform some secure two-party computation.
\item Alice randomly chooses $K_A$ integers in $\{1,\cdots,m\}$, and tells Bob her choices. The chosen set of integers may overlap with the set chosen by Bob.
\item Bob sends the bits $b_j$ and $f_j$ to Alice for the chosen labels $j$.
\item For any chosen label $j$, Alice checks whether $a_j \cdot b_j=e_j \oplus f_j$ holds. If the total number of failures is larger than some preset number of Alice's, she aborts the protocol, or asks Bob to restart the protocol if needed.
\item The remaining one-time tables are regarded as having passed the checking and will be used later in the two-party computing task. They may repeat the steps above to prepare more one-time tables on demand.
\end{enumerate}
\end{algorithm*}

By noting that there is effectively only one bit of classical communication from Alice to Bob in Protocol~\ref{ptl:NLAND}, the analysis for Protocol~\ref{ptl:precompute1} about Bob's data privacy can provide hints for analyzing Alice's data privacy in Protocol~\ref{ptl:precompute1b}. There are analogues of Prop.~\ref{prop1} and Theorem~\ref{thm1} for Alice instead of Bob, see Prop.~\ref{prop2} and Theorem~\ref{thm2} below. To draw an analogy to the analysis of Protocol~\ref{ptl:precompute1}, note that the output bits of Protocol~\ref{ptl:NLAND} can alternatively be written as $r'$ on Alice's side and $(x\cdot y)\oplus r'$ on Bob's side, respectively, where $r'$ is a uniformly random bit. We state the following results. The $I^{\cal M}_x$ is the classical mutual information learnable by Bob about Alice's input $x$ using measurement $\cal M$, in an instance of Protocol~\ref{ptl:NLAND}, where the $\cal M$ incorporates his possible $\cnot$ gate, some Pauli corrections or equivalently some classical postprocessing, and Bell-state measurement with withheld masks. And the other quantities are defined similarly.

\bpp\label{prop2}
In Protocol~\ref{ptl:NLAND}, the follows inequalities hold:
\bea
I^{\cal M}_x+I^{\cal M}_{r'} &\le& 1,\label{eq:info4}\\
I^{\cal M}_x+I^{\cal M}_{x\oplus {r'}} &\le& 1,\label{eq:info5}\\
I^{\cal M}_x+\max(I^{\cal M}_{r'},I^{\cal M}_{x\oplus {r'}}) &\le& 1.\label{eq:info6}
\eea
where the two $\cal M$ are the same in each equation.
\epp

\bpf
The overall communication from Alice to Bob in Protocol~\ref{ptl:NLAND} is effectively only one classical bit, since Alice could apply an arbitrary Pauli operator to the qubit not encoding $x$, while applying a $\sigma_z$ to the qubit encoding $x$ if it is encoded in the $Z$ basis, or a $\sigma_x$ to the qubit encoding $x$ if it is encoded in the $X$ basis. The protocol still works under these changes, with Alice's recording of the value of $t$ changed. Then, if Alice further applies a Pauli operator, the two qubits sent to Bob would be in a maximally mixed state, containing no information for Bob. This shows that the overall communication from Alice to Bob in Protocol~\ref{ptl:NLAND} is effectively only one classical bit. Thus the amount of information that Bob learns about the joint distribution of $x$ and $r'$ is upper bounded by $1$ bit. (As mentioned below, the value of $r'$ is dependent on $t$, so it is not decided by Bob.) The bits $x$ and $r'$ are independent, because $r'$ is an independent uniformly random bit, by the construction of Protocol~\ref{ptl:NLAND}: she takes the XOR of some intermediate result and a uniformly random bit $t$ (generated by herself and independent from $x$) in the last step of Protocol~\ref{ptl:NLAND}. Thus the inequality~\eqref{eq:info4} holds. The bits $x$ and $x\oplus {r'}$ jointly determine $x$ and $r'$, and vice versa, so the amount of information that Bob learns about the joint distribution of $x$ and $x\oplus {r'}$ is upper bounded by $1$ bit. And since the bits $x$ and $x\oplus {r'}$ are independent, we have that the inequality~\eqref{eq:info5} holds. The inequalities \eqref{eq:info4} and \eqref{eq:info5} together imply \eqref{eq:info6}.
\epf

\bt\label{thm2}
In Protocol~\ref{ptl:precompute1b}, Alice's input is asymptotically secure.
\et
\bpf
Similar to the proof of Theorem~\ref{thm1}, we may define the Holevo quantities $\chi_x$, $\chi_{r'}$ and $\chi_{x\oplus r'}$, which are conditioned on the uniform prior distribution for $x$. For the similar reasons as in the proof of Theorem~\ref{thm1}, the following inequality holds for small positive $\e$ and a nonnegative continuous function $g(\e)$,
\bea\label{eq:holevo6}
\chi_x+\max(\chi_{r'},\chi_{x\oplus r'}) \le 1+g(\e),\,\quad\,\notag\\
\mbox{for}\,\,\max(\chi_{r'},\chi_{x\oplus r'})\ge 1-\e,\notag\\
\mbox{where}\,\,g\,\,\mbox{is continuous and}\,\,g(0)=0.
\eea
Note that to show the inequality~\eqref{eq:holevo6} is correct, we need the following implication:
\bea\label{eq:holevoimplication2}
\max(\chi_{r'},\chi_{x\oplus r'})=1 \Longrightarrow \chi_x=0.
\eea
The implication in Eq.~\eqref{eq:holevoimplication2} holds because $\max(\chi_{r'},\chi_{x\oplus r'})=1$ implies that $\max(I^{\cal M}_{r'},I^{\cal M}_{x\oplus r'})=1$ for some measurement $\cal M$ of Bob's, and the latter implies $\chi_x=0$ due to the following argument. Suppose $I^{\cal M}_{r'}=1$ (the case that $I^{\cal M}_{x\oplus r'}=1$ is similar), and consider the four density operators on Bob's side corresponding to four different combinations of $x$ and $r'$,  then the two pairs corresponding to different $r'$ must be orthogonal across the pairs. Then if the states in a pair are partially distinguishable, the left-hand-side of \eqref{eq:info6} would be greater than $1$ for some $\cal M$, which violates Prop~\ref{prop2}. This shows that $\max(I^{\cal M}_{r'},I^{\cal M}_{x\oplus r'})=1$ for some $\cal M$ implies $\chi_x=0$, hence Eq.~\eqref{eq:holevoimplication2} holds.

In the case that Bob's operations are independent among instances of Protocol~\ref{ptl:NLAND}, the security of Alice's input in Protocol~\ref{ptl:precompute1} then follows, for the similar reasons as in the proof of Theorem~\ref{thm1}.

In the following we consider the general case that Bob's operations are not independent among instances of Protocol~\ref{ptl:NLAND}. In such case, the generalization of Eq.~\eqref{eq:holevoimplication2} should hold, due to the similar reason as that after Eq.~\eqref{eq:holevoimplication2}. Then the generalization of Eq.~\eqref{eq:holevo6} for the corresponding Holevo bounds should hold approximately near such extreme point, due to the uniform continuity of the Holevo bounds (as functions of Bob's operations and his messages sent to Alice), and the argument in the proof of Theorem~\ref{thm1} which asserts that the systems in other instances serve as auxiliary systems so the one-copy tradeoff curve of the Holevo quantities still holds. This shows that the argument for the security for the case of independent operations of Bob can be extended to the general case.

For the ``restarts'' of the protocol, the argument is exactly similar to that in the proof of Theorem~\ref{thm1}, so we abbreviate it here.
\epf

It should be noted that when $\chi_x$ is near $0$, there is still some exponentially small probability that Bob may learn quite a significant portion of the information about $x$ in the remaining unchecked instances. The quantitative security level is different from that obtainable by directly adapting Theorem~\ref{thm1} with the roles of two parties switched, at least on the following two points. First, Alice's data privacy has a nonzero lower bound here, see the analysis below Protocol~\ref{ptl:NLAND}. Second, with the same resource cost, Bob's data privacy is somewhat weaker than that in Protocol~\ref{ptl:precompute1}, since some of the one-time tables are used for Alice's checking now. Bob effectively checks about half of the instances as in Protocol~\ref{ptl:precompute1}, and Alice checks the other half. But the security should not be much worse since Bob randomly chooses which instances to check.

In Protocol~\ref{ptl:precompute1b}, if any one party is conservative, his (her) data privacy is guaranteed. But partly due to the possible aborts, it actually suffices to assume either one of the parties is conservative in Protocol~\ref{ptl:precompute1b}, since then the other party might as well be conservative to reach a better security level for himself (herself).

When one party's data privacy is very important, and the other party's data privacy is not too important, we propose the following Protocol~\ref{ptl:precompute2}. It improves the privacy of Alice's input in the later main computation task, while that of Bob's input is somewhat compromised.

\begin{algorithm*}[htb]
\caption{A protocol for generating improved one-time tables with combinations}\label{ptl:precompute2}
\begin{enumerate}
\item Alice and Bob perform Protocol~\ref{ptl:precompute1} or Protocol~\ref{ptl:precompute1b} to obtain some one-time tables after checking. Suppose the instance labeled by $j$ has inputs $a_j$ and $b_j$, and outputs $e_j$ and $f_j$.
\item Bob determines which remaining one-time tables are to be combined into one new instance of one-time table, and tells Alice his decision. Each new instance corresponds to a set $S$ of old instances which satisfy that Bob's input bits are equal (denoted as $b_0$). A new instance has inputs $a'$ and $b'$, and outputs $e'$ and $f'$, where $a':=\sum_{j\in S} a_j \mod 2$, $\,\,b':=b_0$, $\,\,e':=\sum_{j\in S} e_j \mod 2$, $\,\,f':=\sum_{j\in S} f_j \mod 2$.
\end{enumerate}
\end{algorithm*}

In Protocol~\ref{ptl:precompute2}, the privacy of Alice's bit $x$ for the combined one-time table is quite good: The accessible information for Bob is exactly $\frac{1}{2^k}$ bits, where $k$ is the size of $S$ in protocol description. It is because the different one-time tables from the first step are independent. The Holevo bound coincides with the accessible information in the current case.

For the privacy of Bob's input bit $y$ in the combined one-time table, it is possible for a cheating Alice to do a joint measurement on $k$ received states from Bob, to learn the information about $y$ and $r$ simultaneously as much as possible (or $y$ and $y\oplus r$). Bob can deal with this by testing more one-time tables. The resource usage (the amount of entanglement needed and the amount of communication) is estimated to be about $O(t k^2)$ times that of Protocol~\ref{ptl:precompute1}, to achieve the similar level of privacy for Bob, where $t$ is the total number of one-time tables required for the later main computation, and $k$ is the size of $S$ in Protocol~\ref{ptl:precompute2}. In such factor $t k^2$, one $k$ is for the size of $S$, and the additional $tk$ factor means that about $O(tk)$ one-time tables are used in the instance of Protocol~\ref{ptl:precompute1} in the first step of Protocol~\ref{ptl:precompute2}. This factor appears because Alice may use techniques similar to Grover's algorithm to increase the amount of information she may learn about $y$, and the same input variable of Bob's may appear in the original circuit for at most $t$ times. But in the case that the function to be evaluated is for evaluating a program provided by Bob on Alice's data, it is possible that each variable of Bob's  appears only once, then the $t$ factor can be omitted, so that the overhead becomes only $O(k^2)$ compared to the plain use of Protocol~\ref{ptl:precompute1}.

The Protocol \ref{ptl:precompute2} differs from the previous protocols in that it has an extra step of combining the one-time tables, and its usage in the later bipartite computation task may be different by a switch of the roles of Alice and Bob. The success of the quantum protocols is not guaranteed in the presence of cheating, but this does not cause much problem since cheating is caught with high probability, and these protocols are in the preprocessing stage for the overall computation, so the useful data is not leaked. The failures in the quantum gates, measurements, and entanglement generation or qubit transmissions in the preprocessing stage can be tolerated by trial-and-error. The failures in Protocol~\ref{ptl:NLAND} are required to be reported in the protocols, so they have no effect for the testing and later computations. In some experimental implementations the failures might not be reported and might appear as errors, and this would affect the security.

To deal with noise and errors, we propose the following Protocol~\ref{ptl:precompute3}, which has some polynomial resource overhead, and there is a polynomial reduction in error rate in the output.

\begin{algorithm*}[htb]
\caption{A protocol for generating checked one-time tables with reduced error rate}\label{ptl:precompute3}
\begin{enumerate}
\item Alice and Bob perform many instances of Protocol~\ref{ptl:NLAND} to generate some one-time tables, and exchange messages to agree on which instances were successfully implemented experimentally.
\item Alice and Bob uses the method in Protocol~\ref{ptl:precompute1b} to check the instances. They use more instances and a higher ratio of checking by Bob, for better privacy of Bob.
\item Bob chooses a target instance of a one-time table with inputs $a_0$ and $b_0$ and outputs $e_0$ and $f_0$. He also randomly chooses other $q$ instances (called ``auxiliary'' one-time tables) for helping detecting errors in the following step.\label{labelstep3}
\item Suppose the auxiliary one-time table labeled by $j$ has inputs $a_j$ and $b_j$, and outputs $e_j$ and $f_j$, for $j=1,\dots,q$. Bob asks Alice to send $a_0\oplus a_j$ and $e_0\oplus e_j$ to him. For those $j$ with $b_j=b_0$, Bob checks whether $(a_0\oplus a_j)\cdot b_0=e_0\oplus e_j\oplus f_0\oplus f_j$. If there is an error, Bob rejects such target instance of one-time table.
\item The one-time tables which passed the checking will be used later in the two-party computing task. Alice and Bob repeat the steps starting from Step~\ref{labelstep3} to prepare more one-time tables.
\end{enumerate}
\end{algorithm*}

A remark on the method to analyze the security of Protocol~\ref{ptl:precompute3} is as follows. The classical mutual information rather than the Holevo bound is essential in analyzing the security of Protocol~\ref{ptl:precompute3}, which is because the final quantum state of Alice for each instance of the one-time table is measured for performing checkings, while in some other protocols such as Protocol~\ref{ptl:precompute1}, the states for the actually used one-time tables are not measured during the checking.

A strategy for Alice to cheat in Protocol~\ref{ptl:precompute3} is that she could guess the $b_0$ and $f_0$, and confirms her guesses via whether the instance is rejected or not after she sent some bits to Bob for him to check. This way of cheating requires her to also know both $b_j$ and $f_j$, however, she can only guess but not deterministically learn both those bits, because of the checking in Step 2. (As a side remark, Alice cannot take advantage of the fact that some $b_j$ are equal to redesign her attacks in Steps 1 or 2, since the choice of which of them are equal is unknown to Alice.) Hence, such strategy of Alice is impractical. Bob may cheat by partially learning about some $a_j$ through his measurements on quantum states, and then deduce $a_0$ from the sent bit $a_0\oplus a_j$. To deal with this strategy of Bob's, Alice could check for more instances of one-time tables in the initial stage, or combine some target instances, similar to the method in Protocol~\ref{ptl:precompute2}.

It is estimated that the above-mentioned cheating method of Bob requires $q=O({\rm polylog}(\frac{1}{\e}))$, where $\e$ is the target error level in output (assuming the original error level is a constant). This is because if $q$ were larger, Bob would have good chance to cheat while making the outcomes approximately correct. The overall resource costs are not only related to the value of $q$, but also on the number of the one-time tables remaining after Step 2, and the latter is higher than in Protocol~\ref{ptl:precompute1b} because of the presence of errors and the need to detect them. The cost overhead compared to Protocol~\ref{ptl:precompute1b} (in terms of the number of initial one-time tables) is estimated to be $O({\rm poly}(\frac{1}{\e}))$, for Bob's privacy to stay comparable to that in Protocol~\ref{ptl:precompute1b}. The cost overhead compared to Protocol~\ref{ptl:precompute2} should be similar.

From the above, we see that the resource overhead may be exponentially large if the output error rate is required to be exponentially small, while security stays comparable to that in Protocol~\ref{ptl:precompute1b}. A possible explanation is that Protocol~\ref{ptl:precompute3} does not involve active error correction. We suspect that polynomially small error in the output is sometimes acceptable, since the circuit to be evaluated is usually of polynomial length; and some embedded checks in the two-party computation could be adopted so that the computation rewinds to a previous point when some error is detected.

\newpage
\section{Applications in two-party classical computation}\label{sec4}

The following Protocol~\ref{ptl:poly} is for evaluating a linear polynomial with distributed output using the quantum preprocessing protocols introduced above. The linear polynomial is of the form $z=(c+\sum_{j=1}^n a_j b_j) \mod 2$, where $c$ is a constant bit known to Bob, and $a_j$ and $b_j$ are bits on Alice and Bob's side, respectively. The output is the XOR of two bits on different sides.

\begin{algorithm*}[htb]
\caption{A protocol for evaluating classical linear polynomials with distributed output using one-time tables}\label{ptl:poly}
\begin{flushleft}
\noindent{\bf Input:} $n$ bits $a_j$ from Alice, and $n$ bits $b_j$ from Bob, and a bit $c$ known to Bob.\\
\noindent{\bf Output:} a bit $z_A$ on Alice's side and a bit $z_B$ on Bob's side, such that $z:=z_A\oplus z_B=(c+\sum_{j=1}^n a_j b_j) \mod 2$.\\
\begin{enumerate}
\item Alice and Bob perform Protocol~\ref{ptl:precompute1} or~\ref{ptl:precompute1b} or ~\ref{ptl:precompute2} or ~\ref{ptl:precompute3} to obtain $n$ one-time tables.
\item For evaluating the linear polynomial $z=(c+\sum_{j=1}^n a_j b_j) \mod 2$, Alice and Bob perform the evaluation of the nonlocal AND gate for $a_j$ and $b_j$ using the procedure in Sec.~\ref{sec2}, with the output being distributed. They locally calculate the XOR of all bits from the outputs, and Bob additionally takes the XOR with $c$. Each party obtains a bit as the output.
\end{enumerate}
\end{flushleft}
\end{algorithm*}

If Protocol~\ref{ptl:precompute1} is used in Protocol~\ref{ptl:poly}, the data privacy of one party is partial. The leakage is about half of his or her input bits. See also the comments after Protocol~\ref{ptl:generic} below. Generally, we suggest using Protocol~\ref{ptl:precompute1b} or Protocol~\ref{ptl:precompute3} in Protocol~\ref{ptl:poly}, since they at least aim for near-perfect security.

For a generic boolean circuit, we propose Protocol~\ref{ptl:generic}. The main computation after the preprocessing does not include any aborts, and only requires the number of communication rounds to be about equal to the circuit depth. The circuit is assumed to be known to both parties, except for some initial local gates, which may be known only to the local party.

\begin{algorithm*}[htb]
\caption{A protocol for evaluating publicly-known boolean circuits with private bipartite input using one-time tables}\label{ptl:generic}
\begin{enumerate}
\item Alice and Bob decompose the two-party circuit to be evaluated into some local circuits with AND, XOR gates, and some linear polynomials with bipartite input, while adding possible ancillary bits with fixed initial value $1$ (for implementing the NOT gates with the help of XOR). Any nonlocal AND gate in the original circuit is a special case of the linear polynomial.
\item For each AND gate not in the initial stage, the inputs may be distributed, i.e. one or both input bits are the XOR of two remote bits. In the case that both input bits are distributed, Alice and Bob decompose such gate into the XOR of the outputs of two local AND gates and two nonlocal AND gates, the latter being a special case of the linear polynomial. In the case that only one input bit is distributed, they decompose such gate into the XOR of the outputs of a local AND gate and a nonlocal AND gate. For any XOR gate where both input bits are distributed, it is decomposed into two local XOR gates, with the output of the overall gate being distributed. For any XOR gate where only one input bit is distributed, it is effectively one local XOR gate, with the output of the overall gate being distributed.
\item They perform the gates in the resulting circuit in pre-arranged order. The linear polynomials are evaluated using Protocol~\ref{ptl:poly} with distributed output.
\item At the end of the circuit, one party sends some bits to the other party so that the distributed bits for the output are recombined to form the correct output; if there are output on two parties, both parties need to send messages.
\end{enumerate}
\end{algorithm*}

If Protocol~\ref{ptl:precompute1} is used in Protocol~\ref{ptl:generic} with the roles of Alice and Bob switched in the preprocessing only, the data privacy of Bob is partial. The leakage is about half of his input bits in each polynomial. But the privacy is better in the case that the function allows many different inputs of Bob to give rise to the same result. In the case that the function effectively evaluates a universal circuit with data given by Alice and the logical circuit given by Bob, his input has partial privacy which is acceptable due to possible recompilations of Bob's logical circuit. If Protocol~\ref{ptl:precompute2} is used instead of Protocol~\ref{ptl:precompute1}, it is suggested that Alice always be the first party, to save the required number of one-time tables when Alice's data privacy is more important than Bob's data privacy. Then Alice's data in the main computation is asymptotically secure because of the property of Protocol~\ref{ptl:precompute2}. The remarks above are for the noiseless case. For the case with noise, see Sec.~\ref{sec7}, where it is suggested to use Protocol~\ref{ptl:precompute1b} or Protocol~\ref{ptl:precompute3}.

The Protocol~\ref{ptl:generic} has a good property that cheating would usually give rise to wrong results. If some party (partially) cheated in generating some of the one-time tables, so that some but not all of the one-time tables used in Protocol~\ref{ptl:generic} are not secure, then the insecure one-time tables are wrong with some significant probability according to Eq.~\eqref{eq:info3}: the calculation results for a particular nonlocal AND gate would often be incorrect after the distributed output bits are recombined. This implies that the final computation result has large probability to be wrong. But if that party cheated in all the generated one-time tables and passed the other party's test, the computation result could be calculated by the cheating party alone with the help of the messages sent from the other party in the main computation stage. The latter case is not likely to happen, since the other party could set a low threshold in the testing.

Some protocol similar to Protocol~\ref{ptl:generic} could be used for evaluating a public circuit on shared classical secrets between Alice and Bob, when each effective input bit is the XOR of two remote bits. The steps are quite similar except for some initial local gates, so we abbreviate the protocol here.

In the following we discuss the security assumptions. We define Bob to be ``conservative'', if he values the privacy of his input data higher than the possibility to learn Alice's data.

First, let us assume that Bob honestly does the testing in the Protocols~\ref{ptl:precompute1} and \ref{ptl:precompute2}. There could be superpositions in the input and the output of these quantum protocols, but in the later classical computation task, the parties may do computational-basis measurements to force the received superposed states to collapse. Note that one party may insist on using the superposed output from some instance of the one-time table, but when the other party does some later gate using such output as an input, the latter party may do computational-basis measurements to force the collapse of the superposition.

Next, we discuss the case out of the assumption, that is, Bob cheats in the quantum protocols. He may cheat by not aborting after finding that Alice is cheating. This way of cheating is not powerful by itself, but see the following for discussion about his combined ways of cheating. The second way for him to cheat is to use general quantum input (allowing superpositions and entanglement) for the one-time tables, which also allows general quantum output for the one-time tables. In such case, Alice may do computational-basis measurements in the main bipartite computation stage to force the collapse of superpositions. The case that he uses general quantum output for the one-time tables is discussed in the previous paragraph. For the case that Bob combines the two cheating methods above, if Alice is honest, Bob cannot get more information about Alice's data compared to the case of him not cheating in this way. If Alice also cheats, then it is possible that Bob's knowledge about Alice's data on average is better (e.g. when they discard some one-time tables, so that Bob obtains more information about Alice's input in the remaining one-time tables). But that comes at the expense of the higher possible leakage of Bob's data. So a conservative Bob should not do such combined cheating. The third way for Bob to cheat is by using superposed states in the main computation but not the preprocessing. This has no effect since Alice may make a computational-basis measurement on the state received from Bob in the main computation. Note that Alice's data leakage is limited by design of the quantum protocols, except in the case of non-conservative Bob discussed above. In conclusion, if we assume Bob to be conservative, the quantum protocols are asymptotically secure; if we assume Alice to be honest-but-curious, the Protocol~\ref{ptl:precompute2} is asymptotically secure for Alice (as mentioned in Sec.~\ref{sec3}), and in such case it does not make much sense for Bob to cheat since he cannot gain from cheating.

In the following we consider implementing some cryptographic primitives such as 1-out-of-2 oblivious transfer \cite{Crepeau88,Kilian88} and bit commitment. The Protocol~\ref{ptl:ot} is a protocol of 1-out-of-2 oblivious transfer, with its definition shown in the initial part of the protocol. It is constructed based on the one-time table (which effectively implements the PR-box with the help of some communication, c.f.~Sec.~\ref{sec5}) by using the method in \cite{WW05}. In the Rabin oblivious transfer, Alice sends a bit to Bob and it is received with probability $\frac{1}{2}$, and Alice does not know whether the message is received by Bob. The 1-out-of-2 oblivious transfer can be constructed from the Rabin oblivious transfer \cite{Crepeau88}, but we are not aware of a construction for the transformation in the reverse direction, although the oblivious key protocol \cite{WW05}, which is quite similar to the Rabin oblivious transfer, can be constructed from the 1-out-of-2 oblivious transfer according to \cite{WW05}.

\begin{algorithm}[htb]
\caption{A check-based quantum protocol for approximately-secure 1-out-of-2 oblivious transfer}\label{ptl:ot}
\begin{flushleft}
\noindent{\bf Input:} Two bits $m_0$ and $m_1$ on Alice's side, and a bit $b$ on Bob's side.\\
\noindent{\bf Output:} a bit $m_c$ on Bob's side. It is required that Bob does not know $m_{1-c}$, and Alice does not know $b$.\\
\begin{enumerate}
\item Alice and Bob run an instance of Protocol~\ref{ptl:poly} with the polynomial of the form $z=a\cdot b$ (i.e. with only one variable and no constant term), where $a=m_0\oplus m_1$, and $b$ is Bob's input bit. Suppose the output bit on Alice's side is $g$, then the output on Bob's side must be $h=z\oplus g$.
\item Alice sends $m_0 \oplus g$ to Bob. Bob's output bit is $m_0\oplus g \oplus h$.
\end{enumerate}
\end{flushleft}
\end{algorithm}

There are some no-go theorems for quantum bit commitment \cite{LoChau97,Mayers97}. Since our quantum preprocessing protocols allow aborts, and there are some requirements on the players in those protocols, it is still possible that bit commitment can be implemented with the help of the one-time tables generated by the quantum preprocessing protocols. In the Protocol~\ref{ptl:bc} we propose a bit commitment protocol inspired by a computationally-secure construction based on quantum one-way permutations \cite{DMS2000}. Here, instead of using the quantum one-way permutations, we use a special bipartite classical computation with distributed output, with the help of quantum preprocessing. Our scheme is cheat-sensitive and subject to some other assumptions similar to those for the generic Protocol~\ref{ptl:generic}. It requires that one of the parties be conservative.

\begin{algorithm}[htb]
\caption{A check-based quantum protocol for approximately-secure bit commitment}\label{ptl:bc}
\begin{enumerate}
\item Alice and Bob perform Protocol~\ref{ptl:precompute2} to obtain some one-time tables with the degree of security dependent on resource usage. They decide on a large integer $m$ related to the intended security of the current bit commitment protocol.
\item Suppose Alice wants to commit a bit $b$. She asks Bob to together calculate $m$ nonlocal AND gates using the method in Sec.~\ref{sec2}, with her input bits being always $b$, but Bob's inputs are random bits chosen by himself. They obtain some distributed bits as the outcomes. This completes the commit phase.
\item (Reveal phase) Alice sends Bob her output in the instances of the nonlocal AND gates in the previous step. Bob takes the XOR for the corresponding pairs of bits to recover the results of the nonlocal AND gates. From these results, Bob finds out $b$, or decides that Alice has cheated by sending him some random bit string so he cannot recover $b$.
\end{enumerate}
\end{algorithm}

In the last step of Protocol~\ref{ptl:bc}, if Alice sends Bob some random bit string, the results obtained by Bob are generally not consistent with any input value of $b$. For large $m$, it is hard for Alice to guess the appropriate bit string that could make Bob believe the input was $1-b$. The reason is as follows. There are $2^m$ possible bit strings of length $m$ representing the results of the nonlocal AND gates (called ``outcome strings'' below). In the generic case that Bob did not use all zero values for the $m$ input bits for the nonlocal AND gates, one of the outcome strings corresponds to the input value $b$, while a different outcome string corresponds to the input value $1-b$, and all other $2^m-2$ outcome strings are meaningless for Bob. And since Alice does not know Bob's inputs (which we assume to be randomly distributed among the $2^m-1$ nonzero $m$-bit strings) nor his part of the outcome string, she has probability of $\frac{1}{2^m-1}$ of correctly guessing her part of the outcome string corresponding to the input value $1-b$. In the remaining case that Bob had used $m$ inputs bits that are all zero, Alice's input $b$ does not affect the outcome string which is the all-zero string, so Bob cannot distinguish between the case $b=0$ and the case $b=1$, and therefore he should not have chosen such all-zero string as his input. The above analysis means that Bob has an allowed strategy such that a cheating Alice has probability $\frac{1}{2^m-1}$ of success in trying to change the committed bit after making the commit.

\textbf{Coin flipping.} It is mentioned in Appendix H of \cite{NO09} that universally composable oblivious transfer of strings implies coin flipping. And such type of coin flipping is referred to as strong coin flipping in the literature \cite{Mochon05}. The paper \cite{BC91} also gives a protocol for transforming bit commitment into coin flipping. Thus, we obtain that there is a quantum check-based protocol for strong coin flipping.

\section{Applications in two-party quantum computation}\label{sec5}

The methods in this work can be applied in two-party secure quantum computing tasks. When such tasks have classical input and output, they also serve as classical tasks of the type discussed in Sec.~\ref{sec4}, but with quantum implementations. In this way, classical computational tasks are completed with quantum speedup and quantum security advantage. But this requires at least one party to have quantum capabilities beyond those required by Protocol~\ref{ptl:NLAND}. A typical problem in two-party quantum computation is quantum homomorphic encryption (QHE). QHE is an encryption method that allows quantum computation to be performed on one party's private data with the program provided by another party, without revealing much information about the data nor the program to the opposite party.
In this work we present an interactive QHE scheme (``interactive'' means there may be multiple rounds of communication), and a constant-round QHE scheme. The main part of the constant-round scheme has three stages of communication, instead of two in the usual definition of QHE \cite{bj15}. The initial preparation of the one-time tables with checking and preparation of entanglement also involve a constant number of stages of communication.

In the QHE schemes below, there are some polynomials with at least $2n$ variables, where $n$ is the number of qubits in Alice's input. The $2n$ variables correspond to Pauli masks in Alice's teleportation of the input data to Bob. The way Bob changes the coefficients of the polynomials is called coefficient-update rules below. The coefficient-update rules for the first $2n$ variables (and other variables mentioned below) under the action of Clifford gates can be easily obtained from the following relations:
\begin{eqnarray}\label{eq:keyupdate1}
&&\pgate\Xgate=i\Xgate\Zgate\pgate,\quad\quad \pgate\Zgate=\Zgate\pgate,\notag\\
&&\hgate\Xgate=\Zgate\hgate,\quad\quad\quad \hgate\Zgate=\Xgate\hgate,\notag\\
&&\cnot_{12} (\Xgate_1^a \Zgate_1^b \ox \Xgate_2^c \Zgate_2^d)= (\Xgate_1^a \Zgate_1^{b\oplus d} \ox \Xgate_2^{a\oplus c} \Zgate_2^d)\cnot_{12},\notag\\
&&\quad
\end{eqnarray}
where the $\oplus$ is addition modulo 2, and in the gate $\cnot_{12}$, the qubit 1 is the control. The coefficient-update rules for the variables under the $\tgate$ gate can be obtained from the relations
\begin{eqnarray}\label{eq:keyupdate2}
\tgate\Zgate=\Zgate\tgate,\quad\quad \tgate\Xgate=e^{-\pi i/4}\pgate\Xgate\Zgate\tgate.
\end{eqnarray}
The coefficient-update rules are analogous to the key-update rules in \cite{bj15,Dulek16}, but here the coefficients, rather than the Pauli keys (the variables), are updated.

An interactive QHE scheme with almost optimal information-theoretic data privacy and circuit privacy is obtainable by using the method in Protocol~\ref{ptl:poly} to evaluate classical linear polynomials, and using the latter as a subprocedure in the Scheme~4 in \cite{Yu18}. We describe the steps as follows.\\

\textbf{Scheme 1} (An interactive QHE scheme using precomputed one-time tables)
\begin{enumerate}
\item Alice and Bob produce a large number of one-time tables.
\item Alice teleports her $n$ input data qubits to Bob without telling him any Pauli corrections. The $2n$ bits indicating the Pauli corrections are part of the variables in the polynomials to be evaluated.
\item For each stage of the circuit consisting of some Clifford gates and a $\tgate$ gate, the two parties do the following: Bob calculates the coefficients (including the constant term) in the linear polynomial to be used for deciding the $\pgate^\dag$ correction after the $\tgate$ gate. Alice and Bob each does their own part of operations in Protocol~\ref{ptl:poly} to evaluate the current linear polynomial, which has $2n$ variables. This includes the following: each party takes the XOR of the variables (or coefficients) with the input of some unique one-time table, and sends the resulting bits to the other party, and then each party calculates a bit as a part of the distributed outcome of the linear polynomial. According to the local outcome bit, each party does his or her part of the operations in a garden-hose gadget (shown in Appendix~\ref{app2}). The result for a Bell-state measurement corresponding to a $\Xgate^j \Zgate^k$ correction is recorded as two bits $j$ and $k$. The measurement outcomes on Alice's side are part of the variables of the later polynomials.
\item After the last $\tgate$ gate, Bob performs the last Clifford gates in the desired circuit, and calculates his coefficients in the last polynomials for calculating the final Pauli masks. He does his part in evaluating those polynomials, while Alice also does her part. This includes each party sending the XOR of variables (or coefficients) with the local input bit in one-time tables. Each party obtains a bit, and the XOR of these two bits is the intended outcome of the polynomial. Bob teleports his output state to Alice while modifying the correction bits in the teleportation by taking the XOR of those correction bits with his part of the outcomes for the last polynomials.
\item Alice corrects the received state from teleportation with the corresponding Pauli operators, which are determined from Bob's message as well as her part of the output of the last polynomials. The resulting state is the final quantum output.
\end{enumerate}

The following is an estimate of the resource cost of Scheme 1. Suppose $R$ is an upper bound on the number of $\tgate$ gates in the circuit to be evaluated. The number of variables in a linear polynomial is at most $2n+4R$. The factor $4$ is from that each Bell-state measurement has two outcome bits, and Alice has two Bell-state measurements in each gadget. As there are $R+2n$ linear polynomials to be evaluated, and each variable requires a one-time table in the evaluation of a polynomial, the total number of consumed one-time tables is $O(n^2+R^2)$. This is much smaller than the constant-round Scheme 2 below, which has cost exponential in the $\tgate$-gate depth of the circuit. We still introduce Scheme 2 since there are interpolations between the Scheme 1 and the Scheme 2, giving rise to some tradeoff between the number of rounds and the resource cost: the number of rounds may be fewer than in the interactive scheme, while the number of required one-time tables may be higher. This is achieved by running the Scheme 2 for a segment of the circuit, and the two parties interact,  and proceed to the next segment.

The Scheme 2 below is a three-message QHE scheme, with the main structure modified from some scheme with non-ideal security in \cite{Yu18}. A main technique of the scheme is to use a simplified version of a garden-hose gadget from \cite{Dulek16} (and attached in Appendix~\ref{app2}). The main part of the scheme has three stages of classical communication: from Bob to Alice, and from Alice to Bob, and a final teleportation from Bob to Alice. The scheme requires using some linear polynomials of the form in Protocol~\ref{ptl:poly}, but also some nonlinear polynomials, which can also be treated as linear polynomials (with the variables being the product of some original variables) in order to apply Protocol~\ref{ptl:poly}. The construction of the scheme depends on the following property: Bob's coefficients of the (nonlinear) polynomials (the constant term is not included here) do not depend on Alice's original Pauli mask bits or her measurement outcomes in the garden-hose gadgets. The latter independence is possible because we include Alice's previous measurement outcomes and her original Pauli mask bits, as well as her input bit for the garden-hose gadgets as variables. The XOR of Alice's and Bob's inputs for a garden-hose gadget correspond to a polynomial of previous variables, thus Bob's input to any garden-hose gadget can be expressed as a (nonlinear) polynomial of previous variables XORed with Alice's input to this garden-hose gadget, the latter being a new variable. Given his input in the garden-hose gadget, Bob's choice of the pairs of qubits to measure is fixed, independent of Alice's input in the garden-hose gadget. Then the coefficients of Bob in the polynomials can be regarded as independent of Alice's variables. Given the above choice of variables, the constant terms in the polynomials are determined by Bob's local measurement outcomes in his part of the garden-hose gadgets.\\

\textbf{Scheme 2} (A three-message high-cost QHE scheme using precomputed one-time tables)
\begin{enumerate}
\item Alice and Bob each calculates the (nonlinear) polynomials locally according to the circuit to be computed. (The positions of the $\tgate$ gates in the circuit is known to both parties.) They produce a sufficient number of one-time tables.
\item Bob calculates the XOR of each coefficient in the (nonlinear) polynomials with his input in a unique precomputed one-time table, and sends the resulting bits, and the labels for the corresponding one-time tables to Alice.
\item Alice teleports her $n$ input data qubits to Bob without telling him any Pauli corrections. The $2n$ bits indicating the Pauli corrections are part of the variables in the polynomials to be evaluated. With the received message, Alice computes her part of the output of the (nonlinear) polynomials using the one-time tables, based on the method in Protocol~\ref{ptl:poly}. Alice records her part of the output of a (nonlinear) polynomial as a new variable, and according to its value, she does some appropriate $\pgate^\dag$ gate followed by Bell-state measurements in the garden-hose gadgets (shown in Appendix~\ref{app2}). The result for a Bell-state measurement corresponding to a $\Xgate^j \Zgate^k$ correction is recorded as two bits $j$ and $k$. The measurement outcomes are part of the variables of the later polynomials. She calculates the XOR of each term in the next polynomial and her input bit in a unique one-time table, and sends the resulting bits to Bob. She proceeds to do this until she reaches the end of the circuit, including sending messages about the last $2n$ polynomials for the Pauli corrections.
\item Bob receives Alice's message and calculates his output for the first polynomial (which is linear) using Protocol~\ref{ptl:poly}. The Bob part of the output of the first polynomial decides which measurements he should do in the first garden-hose gadget. He performs the Clifford gates and the $\tgate$ gate before the first garden-hose gadget, and performs the appropriate measurements in the first garden-hose gadget. The outcomes of those measurements help determine the constant term in the later (nonlinear) polynomials. He continues to do the next batch of gates and measurements. He evaluates some (nonlinear) polynomial, and according to his part of the output value of such polynomial, he performs the appropriate measurements in the corresponding garden-hose gadget. After the last $\tgate$ gate, Bob does his part in evaluating the last polynomials for calculating the final Pauli masks. The outcomes of those polynomials are distributed as the XOR of bits on the two parties. Bob teleports his output state to Alice while modifying the correction bits in the teleportation by taking the XOR with his part of the outcomes of the last polynomials.
\item Alice corrects the received state from teleportation using the corresponding Pauli gates, which are determined from Bob's message as well as her part of the output of the last polynomials. The resulting state is the final quantum output.
\end{enumerate}

We analyze the resource cost of Scheme 2. The number of variables in the polynomial in the first stage is $2n$, but the polynomial at the second stage (to be evaluated before the second garden-hose gadget) would be nonlinear and has $2n+(2n+1)\times 2+2$ terms apart from the constant term. The term $(2n+1)$ corresponds to Bob's input bit in the first garden-hose gadget, which is not known to Bob before Alice sends her messages, thus such input bit is regarded as a polynomial function of the $2n$ initial Pauli masks and Alice's input bit in the first garden-hose gadget (regarded as a variable). The term $2$ stands for the $\Xgate$ and $\Zgate$ corrections from Alice's measurement outcomes in the garden-hose gadget, where each correction is the XOR of two outcome bits (one in each Bell-state measurement). The final term $2$ is for two of Alice's Pauli corrections in such gadget. They could be the $\Xgate$ and $\Zgate$ corrections for the later pair, or the $\Xgate$ and $\Zgate$ corrections for the first pair. This is enough because the other two corrections were absorbed in the counting above.

Suppose the $k$-th polynomial has $f(k)$ terms apart from the constant term. Then $f(k)=f(k-1)+[f(k-1)+1]\times 2+2$ when $k>1$, so when $k=R$, the number of terms is $O(n\cdot 3^R)$. There are $R$ polynomials (one for each $\tgate$ gate, for evaluating the $\Xgate$ correction before the $\tgate$ gate) which follow the induction rule above. But the last $2n$ polynomials do not follow the rule, and they do not increase any number of variables compared to previous polynomials, because they are for the Pauli corrections after a Clifford circuit. Thus the total number of consumed one-time tables is $O(n^2\cdot 3^R)$.

The security of the Schemes 1 and 2 are optimal if the one-time tables have ideal security, where ``optimal'' means that Alice may learn information about Bob's input from the final output only, and Bob learns nothing at all about Alice's input. But in fact, the one-time tables have partial security, due to the finite number of checks and the noise (including errors), so the security of the schemes above is partial. See also the discussion below.

There are two points on which the security of two-party quantum computation may be somewhat weaker than in classical two-party computation based on the similar procedures for generating one-time tables. First, it is less natural in the quantum protocol to impose classicality of the output of the one-time table. Imposing classicality of course helps security, but it is not necessary given our assumptions about the players in the preprocessing stage. In practice, we may assume that the output of the one-time tables have decohered prior to the use in the main computation. Second, in the schemes given above, the Pauli masks for the original input qubits are used as the variables in all the polynomials involved, this means the data privacy is worse than in the case of classical bipartite computation, in which the intermediate variables replace the roles of the initial variables in many of the linear polynomials. But the use of the quantum preprocessing in this work would give rise to better data privacy than some of the schemes in later parts of \cite{Yu18}, because those schemes require correlated encoding of the different variables, while the variables in the current work are encoded independently by the one-time tables.

We now consider two-party computations in which the circuit is known to Bob only, and each party has some private (quantum) input data. A simple extension of the interactive QHE scheme works, where the extension is just by adding some input qubits on Bob's side. These qubits are not subject to any Pauli masks.

In the following we consider two-party quantum computations with publicly known circuit and private quantum inputs on both parties. One method is to use the simple extension of the interactive QHE scheme as in the last paragraph. A simplified method is to make use of the fact that the circuit is publicly known. We briefly describe it below.

Since the circuit is publicly known, those one-time tables for the linear polynomial for the \emph{first} $\pgate^\dag$ correction after the first $\tgate$ gate are not needed, since Alice can calculate by herself the contributions to this $\pgate^\dag$ correction due to her original Pauli masks. She could just tell Bob before the protocol starts to choose a fixed input on his side in the first garden-hose gadget, then she could decide her input for this gadget on her own. But Bob's measurement outcomes in the garden-hose gadgets are not known to Alice, and they should affect the subsequent $\pgate^\dag$ corrections. Hence, in later garden-hose gadgets, Bob's input cannot be fixed, and the rest of the scheme is similar to the interactive scheme, but with some extra (quantum) input data on Bob's side. In the case that Alice's input is classical, the initial teleportation can be replaced with classical communication with withheld bit-flip masks. If the output is on Bob's side, Bob need not send any message after Alice's message, and Alice sends him some bits for Pauli corrections at the end. In the case that the output is on Alice's side and is classical, the final teleportation from Bob to Alice can be replaced with classical communication without any masks.

\section{Application in check-based implementation of no-signaling correlations with the help of inert communication}\label{sec6}

The Protocol~\ref{ptl:NLAND2} for generating the one-time tables together with Protocol~\ref{ptl:precompute1b} for checking them effectively implement the PR-box (Popescu-Rohrlich box \cite{Popescu1994}) type of correlations. The implementation needs time in communication, and involves some \emph{inert communication}, i.e. sending of some classical messages which do not contain useful information about the inputs (in the ``useful'' one-time tables, but not in those one-time tables subject to checks and not actually used). So this is not a direct implementation of the PR box, which must be instantaneous. Rather, it is a \emph{check-based} implementation of the PR-box type of correlations with time cost and inert communication cost. The fact that it is check-based implies that it is not a deterministic protocol, but \emph{forced almost-deterministic}, meaning that the checking party could set the threshold to very low so that the other party must be nearly completely honest to avoid aborting, and if the parties are indeed nearly completely honest, the protocol is almost deterministic. However, in Protocol~\ref{ptl:NLAND2}, after the initial entanglement has been established, the two directions of teleportation and partial sending of the measurement outcomes can be done simultaneously. This does have some partial flavor of ``instantaneous'' implementation.

In the following, we show how to implement the following general type of no-signaling correlations in \cite{PPK09} in the check-based way.
\bea\label{eq:generalNS}
P(A\oplus B=ab\vert a,b)=\frac{1}{2}(1+E),
\eea
with $0\le E\le 1$. According to an argument in \cite{MAG06} (also mentioned in \cite{PPK09}), the form \eqref{eq:generalNS} is representative of a large class of no-signaing correlations (those with input and output dimensions $2$ on both sides). The way to implement the no-signaling correlations above is similar to the implementation of the PR-box correlations above, but with an additional step in those instances of Protocol~\ref{ptl:NLAND2} not subject to checking but used for the final correlations: Bob randomly flips his output bit with probability $\frac{1}{2}(1-E)$. Such probabilistic step is not involved in the instances of Protocol~\ref{ptl:NLAND2} subject to checking, so Protocol~\ref{ptl:precompute1b} still applies, although with the output correlations changed. A drawback of such implementation is that Bob knows the original value of his output bit, so he may recover a PR-box type of no-signaling correlation. A non-perfect way of dealing with this is to change the last step to that Alice and Bob both flip the respective output bit with some probability $p=\frac{1-\sqrt{E}}{2}$ so that $(1-p)^2+p^2=\frac{1+E}{2}$. Such modified protocol still has the similar drawback that one party could recover a no-signaling correlation with parameter larger than intended.

\section{Discussions}\label{sec7}

{\bf 1. Extensions of protocols}

The qubit-based quantum protocols in this work can be generalized to work for qudits in principle. This is inspired by the classical case in \cite{Beaver98}. This requires some changes in the classical usage of the generated correlations.

The methods in this work are extendable to multipartite classical computation in principle. Some pairs of parties (possibly including some server) may prepare one-time tables using the quantum protocols in this work.

A method of enhancing the security by additional checks after the computation is as follows. If one party, say Alice, does not require the long-term security of her input in the main computation, Bob may ask her to do additional checking of the one-time tables used in the main computation, at a time such that her input data is no longer sensitive, to make sure that she has not cheated by a lot. Of course, in some practical applications, the final computation result provides some check against Alice's cheating, since Alice usually has to cheat all the way to the end for a generic computation to be correct (provided that the final result is on her side, not distributed as the XOR of remote bits), and always cheating successfully is unlikely to happen because of the inequalities in Sec.~\ref{sec3}.

Due to experimental limitations and the overhead from the checkings, the number of one-time tables generated by our quantum protocols may be insufficient if a large two-party computation is to be performed. In that case, it is possible to use some classical processing to achieve a large amount of oblivious transfers (with lowered security) for use in the computation: first turn the quantum-generated one-time tables into the dynamical resource of oblivious transfer using Protocol~\ref{ptl:ot}, and then use the method for oblivious transfer extensions, such as in \cite{Asharov2017}, to generate more oblivious transfers. Note that \cite{Asharov2017} considered 1-out-of-2 oblivious transfers of strings, rather than bits, but the methods should apply to the latter case as well. Since the original oblivious transfers have information-theoretic security while the extensions are computationally secure, we may say that the obtained oblivious transfers have ``mixed'' security.

{\bf 2. Physical implementations of Protocol~\ref{ptl:NLAND}}

The Protocol~\ref{ptl:NLAND2} is an entanglement-based version of Protocol~\ref{ptl:NLAND}. The shared entanglement in Protocol~\ref{ptl:NLAND2} could be prepared by a fixed entanglement-generating device, allowing for failures in preparation (although we allow failures in the whole Protocol~\ref{ptl:NLAND2}, so failures in any particular step is not of much concern). This may also help getting rid of the issue of multiple photons in direct communication, which would harm Bob's data privacy (although some schemes with the direct sending of photons may also allow the detection of multiple photons). Using generation of entanglement could also increase the allowed distance between Alice and Bob, if the entanglement is generated by a device at the middle, compared to using direct sending. The Protocol~\ref{ptl:NLAND3} is also an alternative to Protocol~\ref{ptl:NLAND}, as it uses direct sending in one direction only, with four qubits sent at a time, compared to two qubits in Protocol~\ref{ptl:NLAND}. As for detector inefficiencies and dark counts, the fact that the Protocol~\ref{ptl:NLAND} can be redone after failure can help mitigate the effects of these issues. The appeal of Protocol~\ref{ptl:NLAND} is mainly in that only two qubits are used (although the optical implementation of the $\cnot$ gate with checking for multiple photons might involve some ancillary qubits).

{\bf 3. Effects of noise and errors}

If direct sending of photons is used in Protocol~\ref{ptl:NLAND}, we suggest using the known methods such as decoherence-free subspaces or quantum codes, to reduce or prevent the errors in the transmission. We leave the details for future work. In the following, we analyze the theoretical impact of noise (including errors) on our protocols.

We consider the case that the main computation is classical, since the quantum case is similar in that it also involves evaluating classical linear polynomials. When Protocol~\ref{ptl:precompute1} with noise is used for a bipartite classical computation task, and if Alice's data privacy is more important than Bob's, we suggest that Alice who is the first party in the main computation be the second party in the preprocessing. Then the data leakage of Alice is about the product of the circuit size (the number of the one-time tables) and a small constant indicating the noise level. This is because in Protocol~\ref{ptl:precompute1}, the physical errors and the first party's cheating look about the same for the second party in the verifications (the ``first party'' in this sentence is the Bob in the main computation). For circuits with a high level of parallelism, the data leakage of Alice per input bit is about the product of circuit depth and the error constant described above. So the allowed circuit depth is a constant, which is inverse proportional to the error constant. Similar remarks can be said for Protocol~\ref{ptl:precompute1b} for both sides.

If Protocol~\ref{ptl:precompute2} based on Protocol~\ref{ptl:precompute1} is used for a bipartite classical computation task, we suggest that Alice be the first party both in the preprocessing and the main computation. The noise level is almost not related to the data privacy of Alice, which is exponentially good as the number of one-time tables used in Protocol~\ref{ptl:precompute2} increases. The noise mainly affects the correctness of the computation, and Bob's data privacy. If the noise level is not too low, Bob's data privacy in Protocol~\ref{ptl:precompute2} would not be too good, since he has some identical inputs, and Alice could try to learn partially about each of them to recover his true input. Bob could check more one-time tables to deal with this problem. Thus some polynomial overhead is needed to achieve the similar privacy of Bob's as in Protocol~\ref{ptl:precompute1}. An alternative would be simply using Protocol~\ref{ptl:precompute1b}. The Protocol~\ref{ptl:precompute3} is better than Protocol~\ref{ptl:precompute1b} in the correctness, but it has some overhead in resource costs. A more complicated method is using Protocol~\ref{ptl:precompute2} with ``recompilation'', that is, using some new publicly-known function instead of the original function, with Bob's input changed accordingly, while Alice's input is unchanged, so that the result is the same as the original function with the original input of Bob. If the new function is chosen so that it encodes universal classical circuits, and the possible new inputs of Bob are long enough, we can achieve a good level of security for Bob's input. Such recompilation can be done by classical preprocessing.

There have been studies of the effects of noise in classical cryptographic tasks, and noise is not always bad for security \cite{CK88}. Note that adding some assumptions about quantum capabilities may improve the security in bit commitment \cite{Salvail98}. Adding similar assumptions on top of our quantum preprocessing protocols may improve the security in the applications.\\

\section{Conclusion}\label{sec8}

We have proposed some quantum protocols for approximately generating a certain type of classical correlations (a special case of the one-time tables \cite{Beaver98}) with varying degrees of privacy, to be used in bipartite secure computation tasks. We discussed the effects of noise, and proposed a protocol for dealing with it. We have shown how to use the generated one-time tables in evaluating linear polynomials and generic boolean circuits, and in cheat-sensitive 1-out-of-2 oblivious transfer and cheat-sensitive bit commitment, as well as in (interactive) quantum homomorphic encryption and general two-party secure quantum computation. In the discussions we have mentioned that our method gives a check-based implementation of the PR-box type of correlations, but with some communication time cost, and involves sending of classical messages which do not contain useful information about the inputs, so it is not a direct implementation of the PR box. Some other no-signaling correlations can also be generated in the checked-based way with the help of similar classical communications. Open problems include: applications in check-based quantum implementation of other cryptographic primitives, which may be weaker than the plain version of the primitives; whether there is a constant-round (check-based) information-theoretically secure QHE scheme with costs polynomial in circuit size; a refined analysis of the protocols, taking into account the physical errors in quantum states and operations; fault-tolerance; application to special classes of circuits or functions; design of experimental schemes.

\smallskip
\section*{Acknowledgments}

LY thanks Yingkai Ouyang for helpful comments. This research is funded in part by the NKRDP of China (No. 2016YFA0301802), the National Natural Science Foundation of China (No. 11974096), the Scientific Research Fund of Zhejiang Provincial Education Department (No. Y201737289), and the startup grant of Hangzhou Normal University.

\linespread{1.0}
\bibliographystyle{unsrt}
\bibliography{homo}

\begin{appendix}

\section{An entanglement-based version of Protocol~\ref{ptl:NLAND}}\label{appent}

In this appendix we introduce Protocol~\ref{ptl:NLAND2} which is a variant of Protocol~\ref{ptl:NLAND} based on initial entanglement. It contains only communication from Bob to Alice after the entanglement is established. It does not explicitly contain classical communication from Alice to Bob. But this is because the input $x$ is generated by the measurement in the protocol. If $x$ were generated by Alice before the protocol, one bit of classical communication from Alice to Bob would be needed. For the procedure of testing that the entangled states are indeed EPR pairs, we suggest using a method similar to that using the CHSH inequality in \cite{Ekert91}, which is for testing the singlet state,  but note that we need to leave some EPR pairs untested for later use in our protocol. There are other ways of testing, in which each party measures in one of some different bases, and then the two parties compare notes. These methods generally contain aborts. In Protocol~\ref{ptl:NLAND2}, Bob generates the entanglement, since no explicit communication is from Alice to Bob in the protocol (although Alice's input $x$ implicitly becomes partially known to Bob), so he is less motivated to cheat in entanglement generation.

Note that in studying the security of Protocol~\ref{ptl:NLAND2}, if Alice's (cheating) strategy is such that she does not do any operation (including measurement) on her later two qubits before Bob does anything, Alice's possible ancillary qubits that are involved in her initial operations (if she indeed adds such ancillae) and the possible remaining part of her first two qubits after her initial measurements could be viewed as the purification system for Bob's first two qubits, thus the security analysis of Protocol~\ref{ptl:NLAND} (allowing initial hidden ancillae of Alice's entangled with the sent state) can basically be applied to the analysis of Protocol~\ref{ptl:NLAND2} in such case.

A more complex cheating strategy of Alice is that she does some measurement (including joint measurement on her four qubits) and select only certain outcomes while declaring the instances with other outcomes as ``failed'' to Bob. She could have an advantage in imposing Bob's measurement outcomes, and thus she could learn Bob's output bit $r$ while still learn partial information about $y$, as indicated by numerical calculations. Other strategies could potentially help Alice learn $y$ perfectly but only learn $r$ or $y\oplus r$ partially, but numerical evidence suggests that for any measurement strategy of Alice, the $c$ on the right hand side of Eq.~\eqref{eq:holevo3} cannot be greater than $1.5$. A remedy for such case is that Bob could observe the measurement statistics on his side for the failed instances (or the  instances that did not fail) declared by Alice and his expected statistics of measurement outcomes, to find out if Alice cheated in this way. If Bob finds no deviation from the expected statistics, his reduced density operator of the initial entangled state should be the maximally mixed state on four qubits, and Alice's possible operations can only be local unitaries on her side only, which does not help her learn extra information compared to the original protocol.

Note that in Protocol~\ref{ptl:NLAND}, Alice could potentially have a cheating strategy as follows: she could use some entangled state at the initial sending of qubits, and after Bob's operations, do some (joint) measurement on the ancillae and the returned state from Bob, and declare failure for some measurement outcomes. The security of Protocol~\ref{ptl:NLAND} is more resilient to such attack, since Bob's state is directly sent back. Some simple calculations suggest that the extreme cases in Eq.~\eqref{eq:holevo3} still hold in such case, but the intermediate cases may have somewhat worse security than in the case that Alice did not cheat in this way. As a counter-measure, Bob could check that the average state on two qubits received in the instances that did not fail is the maximally mixed state. Note that such considerations have no effect on the security of Protocol~\ref{ptl:NLAND} when Alice is honest or honest-but-curious. Also note that the Protocol~\ref{ptl:NLAND3} is quite resistent to such attack since there is only one-way communication from Bob to Alice, but again, the intermediate cases need to be studied. But we note that the intermediate cases of the Holevo quantity tradeoffs may become not relevant if Protocol~\ref{ptl:precompute3} is used for checking one-time tables, since the classical mutual information rather than the Holevo bound is essential in analyzing the security of Protocol~\ref{ptl:precompute3}, which is because the final quantum state of Alice for each instance of the one-time table is measured for performing checkings, while in some other protocols such as Protocol~\ref{ptl:precompute1}, the states for the actually used one-time tables are not measured during the checking.

\begin{algorithm*}[htb]
\caption{An entanglement-based quantum protocol for generating one-time tables}\label{ptl:NLAND2}
\begin{flushleft}
\noindent{\bf Input of the generated one-time table:} A random bit $x$ generated in the protocol by Alice, and a random bit $y$ that Bob generates before the protocol.\\
\noindent{\bf Output of the generated one-time table:} $(x\cdot y)\oplus r$ and $r$ on the two sides, where $r$ is a random bit.\\
\noindent The input and output together form the one-time table.\\
\begin{enumerate}
\item After some procedure of generating EPR states and testing them, the two parties share four tested EPR pairs. Bob generates and distributes the entanglement. The testing procedure, which may contain aborts on failure of passing the tests, is discussed in the text.
\item (The steps 3 and 4 performed by Bob can be done concurrently with the Step 2 performed by Alice.) Alice generates a random bit $s$. If $s=0$, she measures the four qubits in her part of the EPR pairs in the $Z$ basis, and records the measurement outcome on the first qubit as $x$; if $s=1$, she measures these four qubits in the $X$ basis, and records her measurement outcome on the second qubit as $x$. The states $\ket{+}$ and $\ket{-}$ are regarded as $0$ and $1$, respectively, in the recording. The XOR of the measurement outcomes on the three remaining qubits is recorded as $g$.
\item If $y=0$, Bob does a $\cnot$ gate on his first two qubits, with the first qubit being the control qubit.
\item Bob teleports his first two qubits to Alice, using the later two EPR pairs, while withholding part of the information about the measurement outcomes: he calculates the XOR of the four correction bits, and sends the resulting bit $w$ to Alice. Bob calculates the XOR of the two bits for $X$ corrections (although they actually correspond to $\sigma_y$ corrections due to the sending of a bit above) on the two qubits, and records the result as his output for the protocol.
\item Alice calculates her output bit: $g\oplus (s\cdot w)$.
\end{enumerate}
\end{flushleft}
\end{algorithm*}

\section{Numerical results for the quantum protocols}\label{app1}

Numerical calculations confirm the inequalities \eqref{eq:info1} through \eqref{eq:info3}.  Note the same $\cal M$ occurs twice in each inequality. The calculations assume that Bob's received a two-qubit mixed state from Alice. This is modeled with a pure state on four qubits, according to the Schmidt decomposition. The calculations assume projective measurements by Alice after she receives the message from Bob, although POVM measurements may give rise to a larger sum on the left-hand-side, and such weakness is remedied by the calculation of the Holevo bound below. Numerical calculations suggest the following inequalities.
\bea
\chi_y+\chi_r \le c,\label{eq:holevo}\\
\chi_y+\chi_{y\oplus r} \le c,\label{eq:holevo2}
\eea
where $c$ is a constant somewhat larger than $1.388$ and is yet to be precisely determined. This implies that
\bea\label{eq:holevo3}
\chi_y+\max(\chi_r,\chi_{y\oplus r}) \le c.
\eea

\begin{figure}[ht]
\centering
\includegraphics[scale=0.33]{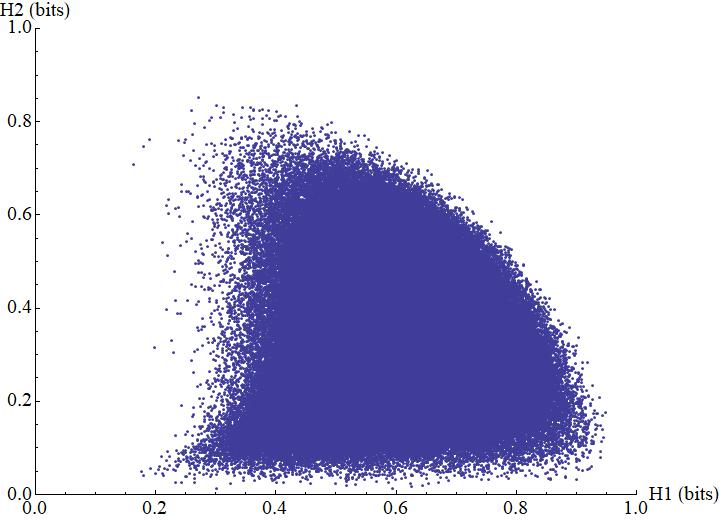}
\caption{An illustration of Eq.~\eqref{eq:holevo3} by numerical calculations. The two axes are the Holevo bounds for $500000$ random mixed states received by Bob from Alice on two qubits, where each mixed state is modeled by a pure state on four qubits including two ancillary qubits. Horizontal axis (H1): $\max(\chi_r,\chi_{y\oplus r})$; vertical axis (H2): $\chi_y$.}
\label{fig:holevo}
\end{figure}

Numerics suggest that near the ends of the tradeoff curve indicated by Eqs.~\ref{eq:holevo} and \ref{eq:holevo2}, one quantity approaches $1$ bit while the other quantity approaches $0$. For some of Bob's received state that approaches the numerically found maximal value of the left-hand-side, the two terms on the left-hand-side of Eq.~\eqref{eq:holevo3} are about equal, and the corresponding sum in the left-hand-side of Eq.~\eqref{eq:info3} under projective measurements is numerically found to be not greater than $1$ bit. The latter sum is observed to have the same property for initial states satisfying $\max(\chi_r,\chi_{y\oplus r})\approx 1$. When there is no ancilla, numerics suggest that the left-hand-side of Eq.~\eqref{eq:holevo3} is not greater than $1$ bit. As quantitative examples for Eq.~\eqref{eq:holevo4}, we have $f(0.1)\approx 0.3$, and $f(0.01)\approx 0.06$. An illustration of Eq.~\eqref{eq:holevo3} by numerical calculations is in Fig.~\ref{fig:holevo}.

\section{The garden-hose gadget that corrects an unwanted $\pgate$ gate}\label{app2}

\begin{figure}[htbp]
\includegraphics[scale=0.455]{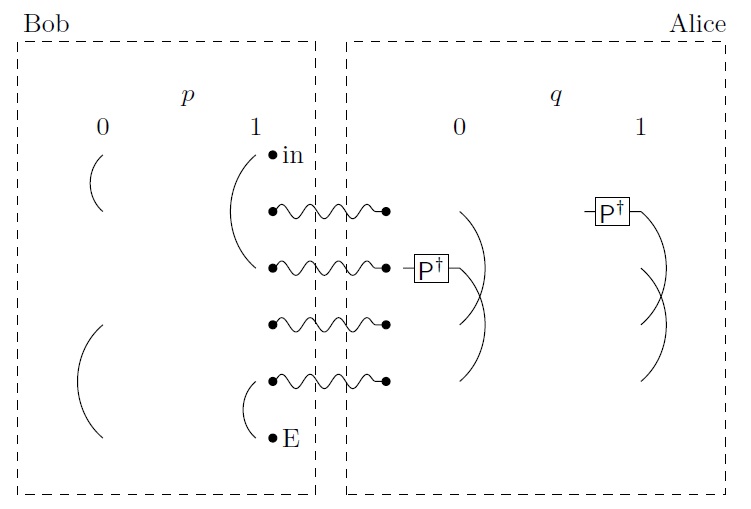}
\caption{A simplified version of a gadget in \cite{Dulek16} for applying a $\pgate^\dag$ to a qubit initially at the position ``in'' if and only if $p+q=1\,\,(\rm{mod}\,\,2)$, using the ``garden hose'' method. The dots connected by wavy lines are EPR pairs. The curved lines are for Bell-state measurements. For example, if $p=0$ and $q=1$, the qubit is teleported through the first and the third EPR pairs, with a $\pgate^\dag$ applied to it by Alice in between. The transformed state of the input qubit always ends up in a qubit on Bob's side which is initially maximally entangled with the qubit labeled ``E'' [in the state $\frac{1}{\sqrt{2}}(\ket{00}+\ket{11})$].}
\label{fig:gadget}
\end{figure}

The Fig.~\ref{fig:gadget} shows a simplified version of a gadget in \cite{Dulek16} for correcting an unwanted $\pgate$ gate due to a $\tgate$ gate in the circuit with certain prior Pauli corrections.

In the figure, the input qubit starts from the position ``in'', and ends up in a qubit which is initially maximally entangled with Bob's qubit labeled ``E'' [in the state $\frac{1}{\sqrt{2}}(\ket{00}+\ket{11})$]. The unwanted $\pgate$ on this qubit is corrected, but some other Pauli corrections are now needed because of the Bell-state measurements. These Pauli corrections are to be accounted for in the later evaluation of polynomials. Note that in each use of this gadget, some of the Bell-state measurements are not actually performed. Alice's two Bell-state measurements are on the same pairs of qubits irrespective of $q$.

\end{appendix}

\end{document}